\documentclass[journal,transmag]{IEEEtran}
\usepackage{newcent}
\usepackage{graphicx, color}
\usepackage{amsmath, amssymb}
\usepackage{subfig}
\usepackage{soul}
\usepackage{url}
\usepackage{booktabs}
\usepackage{multirow}
\usepackage{cite}
\usepackage[linesnumbered]{algorithm2e}

\DeclareMathOperator*{\argmax}{arg\,max}

\begin{document}


\title{On the IP Traffic Matrix Problem\\ in Hybrid SDN/OSPF Networks}


\author{\IEEEauthorblockN{Marcel Caria and Admela Jukan}
\thanks{M. Caria and A. Jukan are with the Technische Universit\"at Carolo-Wilhelmina zu Braunschweig, 38106 Braunschweig, Germany, e-mail:  \{m.caria, a.jukan\}@tu-bs.de}
}


%


\maketitle

\begin{abstract}
IP networks with a hybrid control plane deploy a distributed routing protocol like OSPF and the centralized paradigm of SDN in parallel. The advantages and disadvantages of a hybrid control plane have already been extensively discussed in the literature. This paper examines to what extent hybrid SDN/OSPF can solve the IP traffic matrix and related monitoring problems, inherent to the IP layer. The problem persists in hybrid networks, as the required SDN byte counters for a complete traffic matrix may not be sufficiently implemented (or even not at all), or the deployed SDN nodes may be too few, or not adequately located in the network. For such cases, we propose to augment the SDN traffic statistics with SNMP-based measurements on IP backup links. We address implementation and network function virtualization aspects of the required hybrid monitoring infrastructure and discuss the timing issues of the measurements based on hands-on experiences in our lab. We furthermore provide a placement algorithm for SDN nodes and backup links that can guarantee a complete IP traffic matrix.
\end{abstract}


\section{Introduction}\label{Introduction-section}
\IEEEPARstart{T}{he} IP traffic matrix determines the amount of traffic transferred per second between any ingress-egress pair of routers. It is essential for IP network operation and management, including tasks like traffic engineering, routing protocol configuration, security and reliability, capacity planning, and fault diagnosis. The traffic matrix is however not readily available in legacy IP networks. The measuring of all flows directly is not practical, as it requires a significant amount of monitoring equipment and network-wide configuration efforts~\cite{Papagiannaki}. Therefore, the traffic matrix is usually \emph{estimated} or \emph{sampled}, both leading to inaccurate results that may adversely impact network operations, due to faulty configurations based on uncertain traffic statistics.

Throughput statistics on a per-link basis (commonly referred to as \emph{link loads}) are easily available from all network nodes via Simple Network Management Protocol (SNMP) requests and are typically monitored by network operators. The mathematical relation of the three parameters link load, routing, and traffic matrix is $L=R\cdot F$, where the link load of the $n$ links in the network is the (given) column vector $L = (l_1, l_2, \ldots, l_n)^T$, the demand of all $m$ ingress-egress (IE) flows is the (sought) row vector $F = (f_1, f_2, \ldots , f_m)$, and the routing is represented by the (given) binary $n\times m$ matrix $R$, where $r_{ij}$ is 1 if flow $j$ is routed via link $i$, and 0 otherwise. To put it more intuitively, the load on a link is the sum of all the IE flows that traverse it. Due to the fact that link loads represent \emph{aggregated} flows with their number being (depending on the topology) easily an order of magnitude below the number of IE flows, an attempt to solve the above linear system for $F$ results in a heavily under-determined linear system. The \emph{estimation} of the traffic matrix based on this statistical data means the search for a \emph{good} solution of the described problem, which however typically exhibits severe estimation errors~\cite{Nucci}.

Traditionally, network operators used to deal with the problem either by significant over-provisioning of network resources (which renders the exact knowledge of the traffic matrix unnecessary), or by installing expensive monitoring equipment. However, the advent of Software-Defined Networking (SDN) involves a new and powerful mechanism for network monitoring, as SDN-enabled devices provide additional byte counters for all individual entries in their forwarding tables. We assume that this mechanism will solve the traffic matrix problem once and for all in the long run. Today, however, IP networks are likely to be implemented with a hybrid control plane deploying a distributed routing protocol like OSPF and the centralized paradigm of SDN in parallel. In hybrid networks, the known and difficult problem to generate the IP traffic matrix persists, as the required SDN byte counters may not be sufficiently implemented (or even not at all~\cite{hardware_counters}), or the deployed SDN nodes may be too few, or not adequately located in the network.

This paper examines a new approach to solve the IP traffic matrix and the related monitoring problem in networks with a hybrid SDN/OSPF control plane. We propose to augment the SDN-based traffic statistics with SNMP-based throughput measurements of the absent flows, obtained by temporarily offloading them on IP backup links. We address implementation aspects of the required hybrid monitoring infrastructure and discuss the timing issues of the measurements based on hands-on experiences in our lab. We explain the technical background and possible methodical pitfalls, outline the design of our framework, discuss our practical experiences with OpenFlow- and link-based measurements in our testbed, and address the conformity of our ideas with the novel paradigms in network management such as Network Function Virtualization (NFV). We finally provide an ILP model and a greedy heuristic to determine the optimal measurement locations for SDN nodes and backup links, and we show in our performance evaluation that there is a near linear trade-off between both resources. We conclude from our results that a hybrid control plane with only a few SDN nodes can provide the complete traffic matrix in case multiple backup links are available for measurements.

The rest of the paper is organized as follows: Section~\ref{relatedwork-section} discusses the related work and our contribution. Section~\ref{arch-section} presents the assumed IP network architecture, and Section~\ref{implementation} explains our per-flow measurement technique, implementation and virtualization aspects, and our practical measurement experiences in a testbed. We furthermore present an analytical model for backup link and SDN node location optimization and a fast heuristic in Section~\ref{math-section}. Finally, Section~\ref{performance-section} presents the performance study and Section~\ref{conclusion-section} concludes the paper.


\section{Related work and our contribution }\label{relatedwork-section}

\subsection{Traffic Matrix Estimation}
There are multiple traffic matrix estimation techniques that use link loads and routing information, generally referred to as \emph{network tomography}. Since the related linear system is ill-posed (and thus has multiple solutions), the accuracy of network tomography methods differ based on the statistical assumptions they make. One typical approach is to assume a certain traffic distribution function. Another method is to derive a solution with higher order statistics of the link loads, linear programming, or quadratic programming~\cite{tomography}. The \emph{gravity model}~\cite{gravity} is another traffic matrix estimation technique initially developed for the research on road traffic. Here, the traffic matrix is derived only from the total traffic entering the network at each ingress and the total traffic exiting the network at each egress, whereas the interior network links and routing information are not considered. The gravity model can be used as input to the tomography method, which has been coined as \emph{tomogravity model}~\cite{tomogravity}. Interested readers are referred to~\cite{comparison} for a detailed comparison of different traffic matrix estimation methods for legacy networks. We note that regardless of the method, all proposed traffic matrix estimations typically exhibit average errors in the range of 10\% to 25\% with some flow estimate errors above 100\%~\cite{Nucci}.

Another method to to obtain additional measurements is to periodically reconfigure the routing in the network. Paper~\cite{Nucci} proposes rerouting by altering the IP routing protocol's link metrics in order to create an additional linear system  $L = R \cdot F$ (containing different $R$ and $L$). This new linear system can be combined with the original one to increase the rank. This method is performed repeatedly until the desired rank is achieved. The authors in~\cite{MeasuRouting} propose to route flows over fixed network monitor nodes. Please note that our approach does not require to alter the routing.

The adoption of SDN introduces additional traffic statistics that can be used to improve the estimation of the traffic matrix. \cite{ZhimingHu} proposes to use the SDN-based measurements in addition to link counters to increase the rank of the estimation problem in data center networks. However, despite the assumption of a complete SDN deployment, the paper reasons that measuring every flow in the network is too costly. Consequently, a large-scale flow aggregation for the flow tables maybe required, which in turn results in a yet (not so) underconstrained linear system. Paper \cite{hybridSDN_TM_estimation} provides for the same purpose two efficient algorithms to determine measurement rules, but for hybrid SDN networks, assuming that TCAMs in the SDN switches do not suffice for all required monitoring actions. Contrary to our approach, the current papers do not  attempt to optimize SDN node placement to improve traffic monitoring.

\subsection{Direct Measurements Revisited}
The deployment of IP links between \emph{all} ingress and egress router pairs (a so called full mesh topology) would allow the measurement of the complete traffic matrix only with SDNM-based link loads, which however does not scale, as the number of links would increase quadratically with the number of nodes. To address this issue, other standard layer~2 frameworks can be deployed for this purpose. For instance, MPLS could use LSPs, PBB-TE can provision E-Lines; even with the traditional Ethernet, VLANs could be configured and used for direct measurements of IE flows. Let us consider the case of MPLS as an example; here, the operator can set up an LSP between a pair of routers and install a packet counter on that LSP. In a network with $N$ nodes, this would require the setup of $N\cdot (N-1)$ LSPs, and each LSP setup would involve the configuration of all routers along its routing path. In addition to this significant configuration and management overhead, there is an issue of how widely spread the mentioned frameworks are. For instance, MPLS -- despite its maturity -- is used in only 7\% of all autonomous systems in the Internet and packet forwarding in the Internet backbone is still primarily based on pure IP~\cite{MPLS}. Likewise with E-Lines and VLANs, the configuration overhead remains a concern. 

The problem of choosing the best nodes to perform flow monitoring has already been studied in the context of Cisco's IOS NetFlow feature. This traffic sampling method has impact on the CPU load in the router, which can be significant~\cite{netflow}, and NetFlow must be available on the routers. This is not generally the case, especially in carrier networks, where multiple vendors' equipment is used. The authors of~\cite{monitor} point out that the accuracy of traffic analysis based on flow measurements depend on the sampling rate and the number and placement of monitors, and present methods to jointly optimize the problem. 

A straightforward approach to measure the entire traffic matrix is to monitor traffic on all ingress routers' monitoring ports with cheap off-the-shelf host. A central server can then collect the data along with the routing information from all routers. This solution does however not scale to current transmission speeds in core networks with the typical 100~Gbit/s ports. There are systems on the market that do scale to core network dimensions, like HP's OpenView Dynamic Netvalue Analyzer~\cite{hp-dna}. However, such systems have to be purchased and maintained, and due to their involved high capital and operational expenditures, over-provisioning of network capacity till the point where having exact knowledge of the traffic matrix becomes unnecessary is still considered as the easiest and most cost effective solution by the majority of network operators.

In the case OpenFlow routers are deployed in the IP network, it is indeed possible to measure an IE flow directly on the ingress router. The network operator can identify all flow table entries for a specific egress router at the ingress router and use the according byte counters for this monitoring purpose. However, the required throughput statistics on a per-flow basis may be implemented insufficiently or even not at all~\cite{hardware_counters}.

\subsection{Our Contribution}

In this paper, we propose to use a separate physical port on a pair of IP routers to be configured as a backup to an IP link. A backup link in addition to a regular IP link is easy to create and to configure, while it also allows measurements using regular SNMP link byte counters, which is vendor-independent and available in every router. In contrast to sampling, our method \emph{directly measures} the IE flows using the SNMP link count on the backup link, so that an extrapolation from samples is not necessary. Additionally, we provide a solution for networks during the upgrade to SDN, when there are too few SDN nodes deployed, or they may be insufficiently located in the network.

This work is based on our past work where we studied various aspects of hybrid SDN/OSPF networking, such as technology migration strategies~\cite{migration1, migration2}, network optimizations like capacity planning, traffic engineering, fault recovery~\cite{tnsm}, and fault tolerance~\cite{globecom2016}. We have proposed to combine hybrid SDN/OSPF networking with dynamic optical circuits in~\cite{hpsr}. We have also proposed a novel mode of operation for hybrid SDN/OSPF networks in~\cite{im2015}, where the legacy routing domain is partitioned into subdomains, which allows to some degree to steer the legacy protocol by the central SDN controller. We proposed in~\cite{ManFI} to establish optical bypasses to measure IE flows. We have shown in that work that a relatively small number of strategically placed bypasses in the network allows the measurement of a very high number of flows. At the same time, we observed that an inappropriate high number of bypasses is necessary to gain the full traffic matrix. In this paper, we extend our approach to any form of backup link in the IP layer and additionally use OpenFlow byte counters from all SDN-enabled devices assuming a hybrid SDN/OSPF network architecture, which is novel.

\begin{figure}[t] \center \includegraphics[width=\columnwidth]{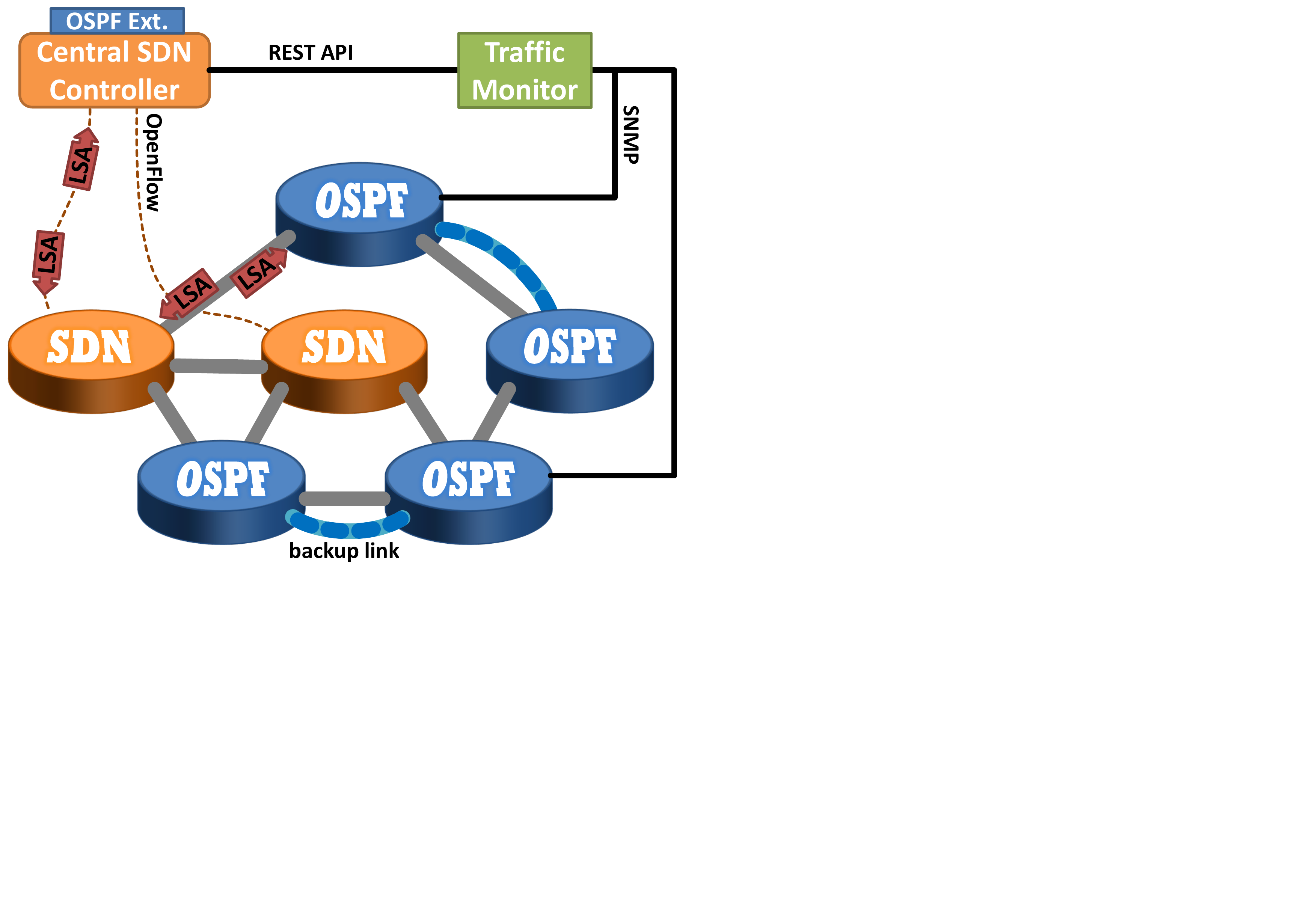}
\caption{Reference network architecture with SDN/OSPF control plane and the proposed backup links for SNMP-based measurements.} \label{arch} \end{figure}

\section{Hybrid Network Architecture}\label{arch-section}
Figure~\ref{arch} shows the reference network architecture, deploying two kinds of routers: legacy OSPF routers (shown as the blue nodes) and OpenFlow-enabled SDN routers (shown as the orange nodes). In SDN networks, a central controller configures the overall network behavior. This paradigm is contrary to the legacy IP network control plane, where distributed protocols (i.e. OSPF or IS-IS) are used, and each network node is configured by its internal controller based on topological information gathered from routing protocol update packets. These packets are referred to as Link State Advertisements (LSA) and they are distributed through the entire routing domain with a protocol-inherent flooding mechanism. SDN, in contrast, allows routers to become simple packet forwarding machines, while all intelligence is implemented in a (logically) \emph{centralized} controller. We assume that the OpenFlow protocol is used, which is the de facto standard for the communication between the central controller and SDN routers. Each SDN device establishes an individual OpenFlow channel to the central controller, depicted as the orange dashed lines in the figure.

As OSPF and SDN routers are not inherently compatible, a network that uses both types of routers requires some sort of hybrid SDN/OSPF control plane. There are different approaches to inter-operate OSPF and SDN, and the simplest version uses backward compatible (i.e. OSPF-enabled) SDN routers like in~\cite{hybridArchPaper}: Here, SDN routers are able to autonomously communicate OSPF-conform with neighboring legacy routers. Such hybrid routers generate their basic forwarding tables based on OSPF, whereas higher priority rules can be installed by the central SDN controller. However, we assume a different version of hybrid SDN/OSPF, where SDN routers are not required to be OSPF-enabled, because all routing protocol processing is performed by the central controller. These SDN routers are simply configured to forward all LSAs to the SDN controller over the OpenFlow channel, and the central controller processes the protocol (possibly with some OSPF extension, like shown in the figure, as common open-source controllers do not support legacy routing) and sends according response packets back to the SDN router (which in turn forwards it to the originating OSPF router). A proof-of-concept implementation of such a hybrid SDN/OSPF control plane was demonstrated in~\cite{routeflow}. Please note that, in contrary to the control plane, there are no incompatibilities between OSPF and SDN in the data plane of the IP layer, as IP packet processing depends solely on a router's individual forwarding information base. 

Our proposed monitoring framework is depicted as the green box in Figure~\ref{arch}. The system requests byte counters from the central SDN controller (via the controller's REST-based northbound API) and from regular and backup IP links (via SNMP). The IP layer's control plane has to deal with backup links to keep them unused under normal network operation, which requires the assignment of an OSPF link cost to the backup link that is larger than the one of the original IP link. Explicit forwarding rules are then created at the ingress router of the backup link to offload specific IE flows onto it. We only consider the measurement of IE flows that already traverse the said routers in their original IP route. These flows are separately rerouted by means of consecutive routing policies at the backup link ingress router, thus there is no impact on regular routing.

For every IE flow that we aim to measure on a backup link, we need to define an access control list (ACL) such that the IE flow in question can be distinguished from the remaining traffic. Please note that ACLs may become complex due to the fact that ingress/egress routers generally handle a lot of network prefixes. We afterwards configure a routing policy at the ingress router that appoints the backup link's port for all packets matching the ACL. In this way we provide that only packets of that specific IE flow are transmitted over the backup link, such that the SNMP link count function can be used for the measurement. We always use only one routing policy at a time per backup link, which limits the overhead in packet processing. In case a highly utilized link is bypassed for measurements, the use of the backup link itself may affect the traffic volume of the re-routed IE flow due to the \emph{elasticity} of TCP connections' throughput. We therefore limit the study to cases of moderately loaded links always safely below the risk of congestion, which is a fair assumption in IP backbone networks.

For our SNMP-based measurement scheme to work, it should be noted that the here considered backup links need to be visible in the IP layer. At the same time, it is known that other backup schemes are common in backbone IP networks, especially those based on layer~2, which provide sub-50ms protection, but are completely transparent to layer~3. An exception from these L2 protection schemes is Ethernet Link Aggregation (IEEE 802.3ad) that allow yet again for SNMP-based measurements, in case its implemented with proprietary aggregation schemes like inactive failover ports or VLAN-based Quality-of-Service mechanisms. All it basically requires for our scheme to work is the possibility to temporarily separate a flow on an otherwise unused port and an open configuration interface.

\begin{figure*}[t] \center
\includegraphics[width=\textwidth]{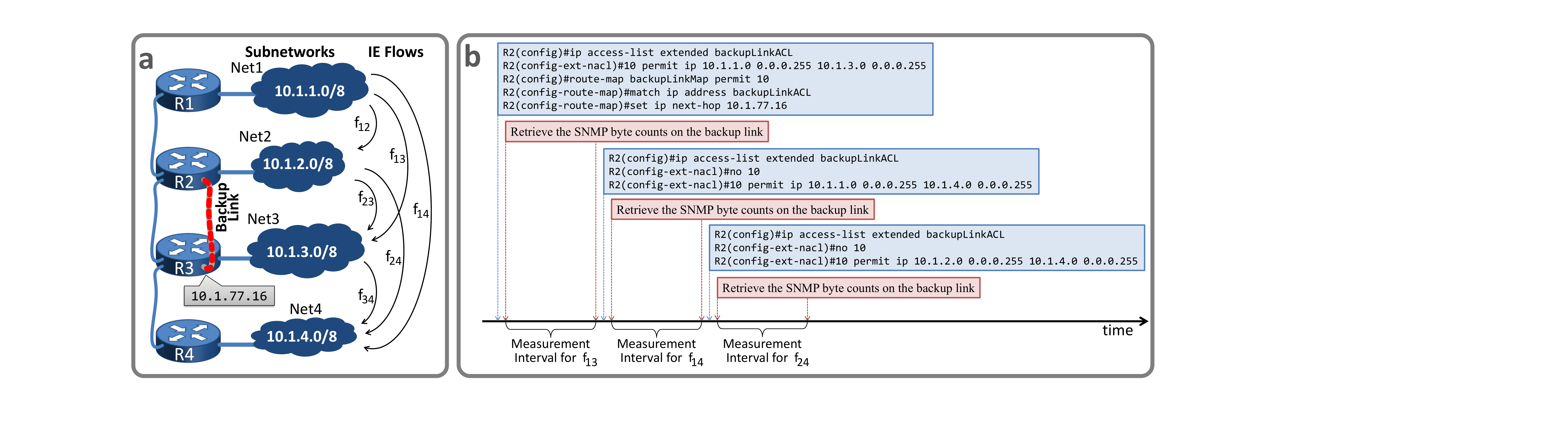}
\caption{Example for the topology (Subfig. a) and configuration (Subfig. b) of a simple backup link setup: the throughput of the flows $f_{13}$, $f_{14}$, and $f_{24}$ are measured on the backup link parallel to link R2-R3.}
\label{example} \end{figure*}

\section{Combining OpenFlow- and SNMP-Based Measurements}\label{implementation}
The throughput of a link is determined based on consecutive queries of its byte counter, which is accessible with SNMP. This protocol is one of the most prevalent network management standards, developed by the IETF to allow the configuration and monitoring of network elements. It is defined following a manager/agent principle, where an SNMP-enabled network element (a router) implements an agent that can configure and monitor the network element and communicate with an SNMP manager. The protocol uses a hierarchical data structure called Management Information Base (MIB), which defines the syntax and semantics of the stored data. Its values ​​are referred to as the MIB objects, and each object has a unique object identifier (OID). Routers commonly provide a MIB object for each port that provides counters for the incoming and outgoing bytes. The throughput of a link can then simply be calculated as the difference between two consecutive byte counter values divided by the time difference between the two queries.

Figure~\ref{example}a shows a 4-node network to illustrate measurements on a backup link. The topology includes four routers, R1 to R4, and for the sake of simplicity, we only consider the six IE flows into the downward direction. The corresponding ill-posed linear system for the traffic matrix of this network has accordingly three rows (from the three original links) and six columns (from the six IE flows). The most beneficial backup link to use is obviously between R2 and R3, since the original IP link R2-R3 carries the most IE flows ($f_{13}$, $f_{14}$, $f_{23}$, and $f_{24}$). As in our example topology the rank of the linear system needs to be increased by three, it is sufficient here to measure only three of the possible four IE flows to let us solve the linear system of the traffic matrix. In other words, if we measure $f_{13}$, $f_{14}$, and $f_{24}$ on the backup link and subtract the sum of their throughput from the load of the original link R2-R3, we obtain the throughput of $f_{23}$.

Figure~\ref{example}b shows the basic configuration steps exemplarily for Cisco IOS (in the blue boxes) and the timing of the corresponding retrievals of the SNMP byte counters from the backup link (red boxes): Assuming that the OSPF metric of the backup link is already configured to be larger than the one of the original link, we create an ACL with the name {\tt backupLinkACL} on the backup link's ingress router R2. This ACL filters all packets between the subnetworks Net1 and Net3. Afterwards, we create a routing policy that forwards all those packets to the backup link. We now have rerouted $f_{13}$ onto the backup link (that doesn't carry any other traffic), which allows us to measure it with two sequent retrievals of the SNMP byte counter on the backup link's port. In the second step, we reconfigure the ACL by deleting its previous matching rule and by adding a new one for flow $f_{14}$ and measure its throughput on the backup link. Then, we similarly configure and measure $f_{24}$ on the backup link. After all measurements are performed, the backup link could be decommissioned (if the layer~2 control plane supports port configuration) and the routing policy and the ACL can be deleted from R2.

The OpenFlow standard defines byte and packet counters for the entries of the flow table, which can be retrieved by the SDN controller. The effort required to retrieve the byte counters from OpenFlow devices in the network is comparably low, as all popular controller implementations provide some (REST-based) northbound interface for management purposes. The throughput of a flow can then be determined -- similar to a link's throughput -- based on consecutive queries of its byte counter in the traversed OpenFlow router.

\begin{figure}[tb] \center
\includegraphics[width=\columnwidth]{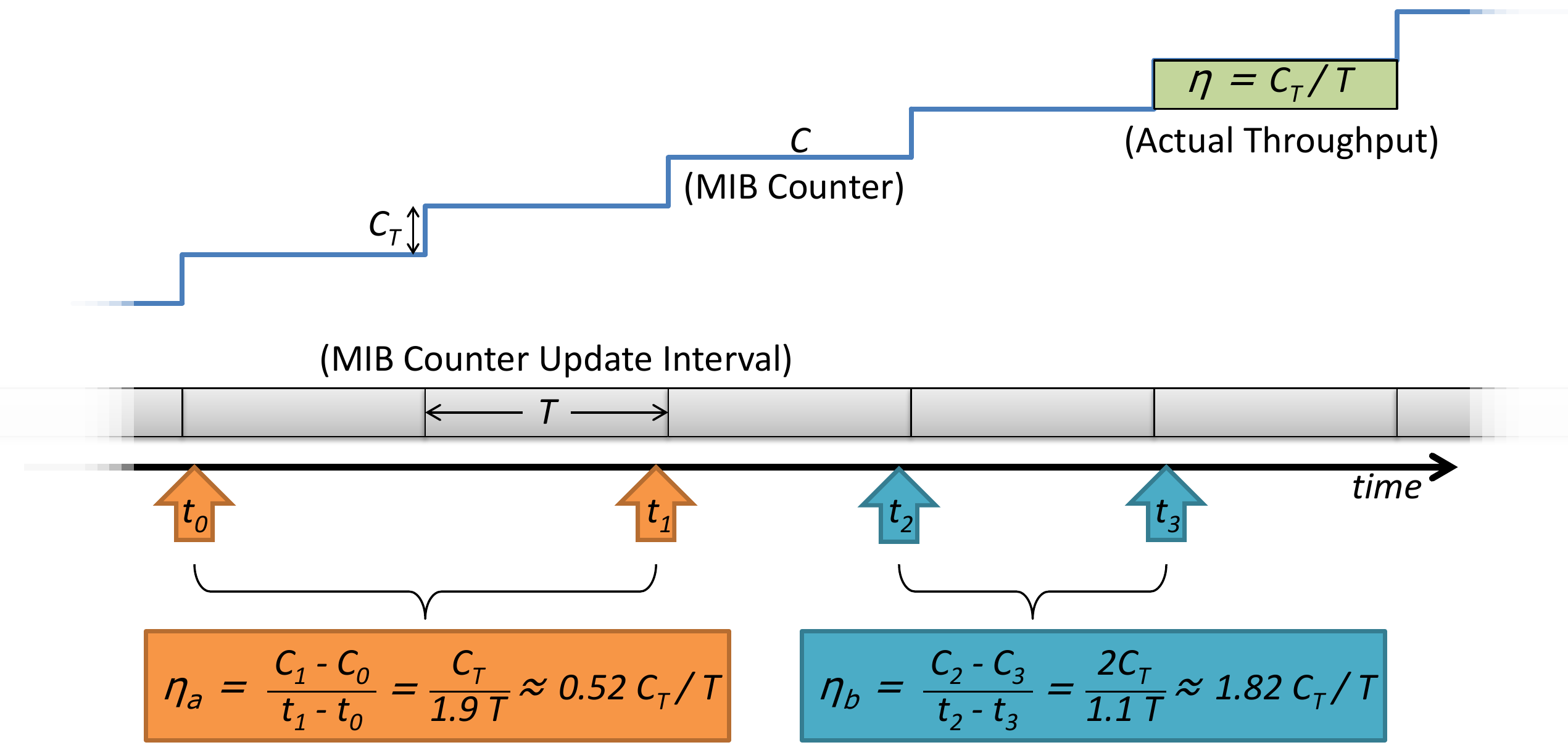}
\caption{Illustration of the MIB update problem}
\label{mib_update} \end{figure}

\subsection{Discussion on Implementation}
Link byte counters are hardware implemented and not directly accessible from the outside. In order to provide that information via SNMP, the counter value is frequently written into the MIB, from where it can be requested. Apparently, the MIB is not synchronously updated with every tick of the counter (at least in the devices tested in our lab), but there is a fixed update interval $P_{MIB}$. Consequently, the calculation of the throughput of a link from its byte counter in the MIB is only straightforward when the measurement interval is long enough: Two counter values $C_1$ and $C_2$ have to be retrieved, and the time instants $t_1$ and $t_2$ at the retrievals have to be known. The throughput can then simply be calculated as
\[ \eta = \frac{C_2 - C_1}{t_2 - t_1} \]
However, this method requires that $t_2-t_1 >> P_{MIB}$, i.e., the measurement interval has to be much greater than the MIB update interval. If this is not the case, the timing of SNMP requests must be synchronized with $P_{MIB}$. Figure~\ref{mib_update} shows two extreme examples in which the calculated throughput differs from the actual throughput.

\begin{figure}[tb] \center
\includegraphics[width=\linewidth]{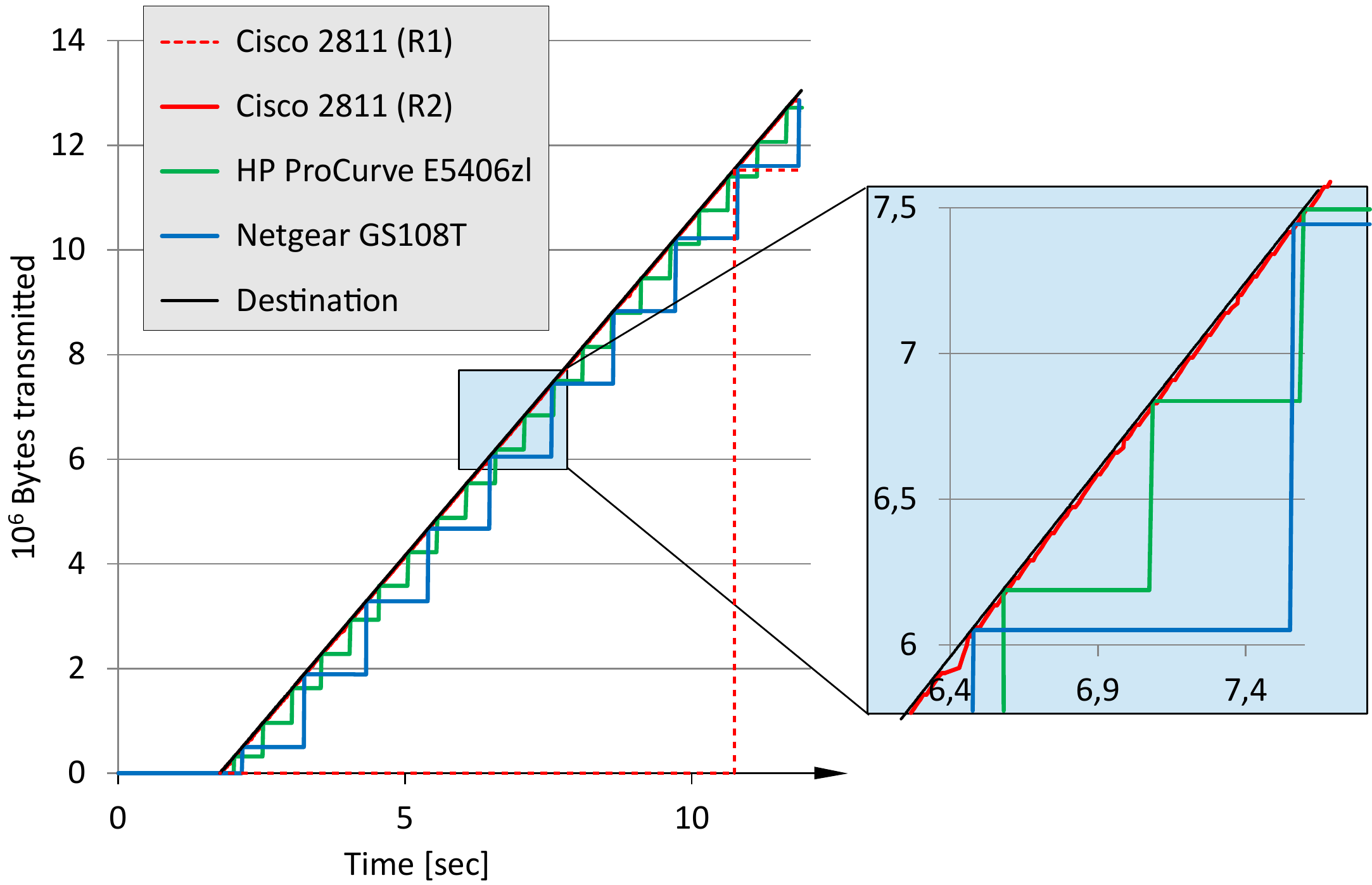}
\caption{The MIB update rates of the different devices measured in our testbed.}
\label{mib_update_rate} \end{figure}

Figure~\ref{mib_update_rate} shows the MIB update rate (for a static throughput) of the different devices used in our testbed, which are: 1)~an HP ProCurve E5406zl switch, 2)~a Netgear GS108T switch, and 3)~a Cisco 2811 router. The number of bytes transmitted over the network were recorded with Wireshark at the destination. The byte counters were requested simultaneously every 10ms from all network devices using SNMP. The MIB update frequencies are revealed by the step functions of the different counters. The HP switch updates every 500ms and the Netgear switch every 1000ms. The Cisco router updates the MIB by default only every 10 seconds (shown as Cisco R1 in the figure), but it can be configured (shown as Cisco R2) to do it at max every 10ms (using the not documented IOS command {\tt snmp-server hc poll $<$interval$>$}).

\begin{figure}[b] \center
\includegraphics[width=\linewidth]{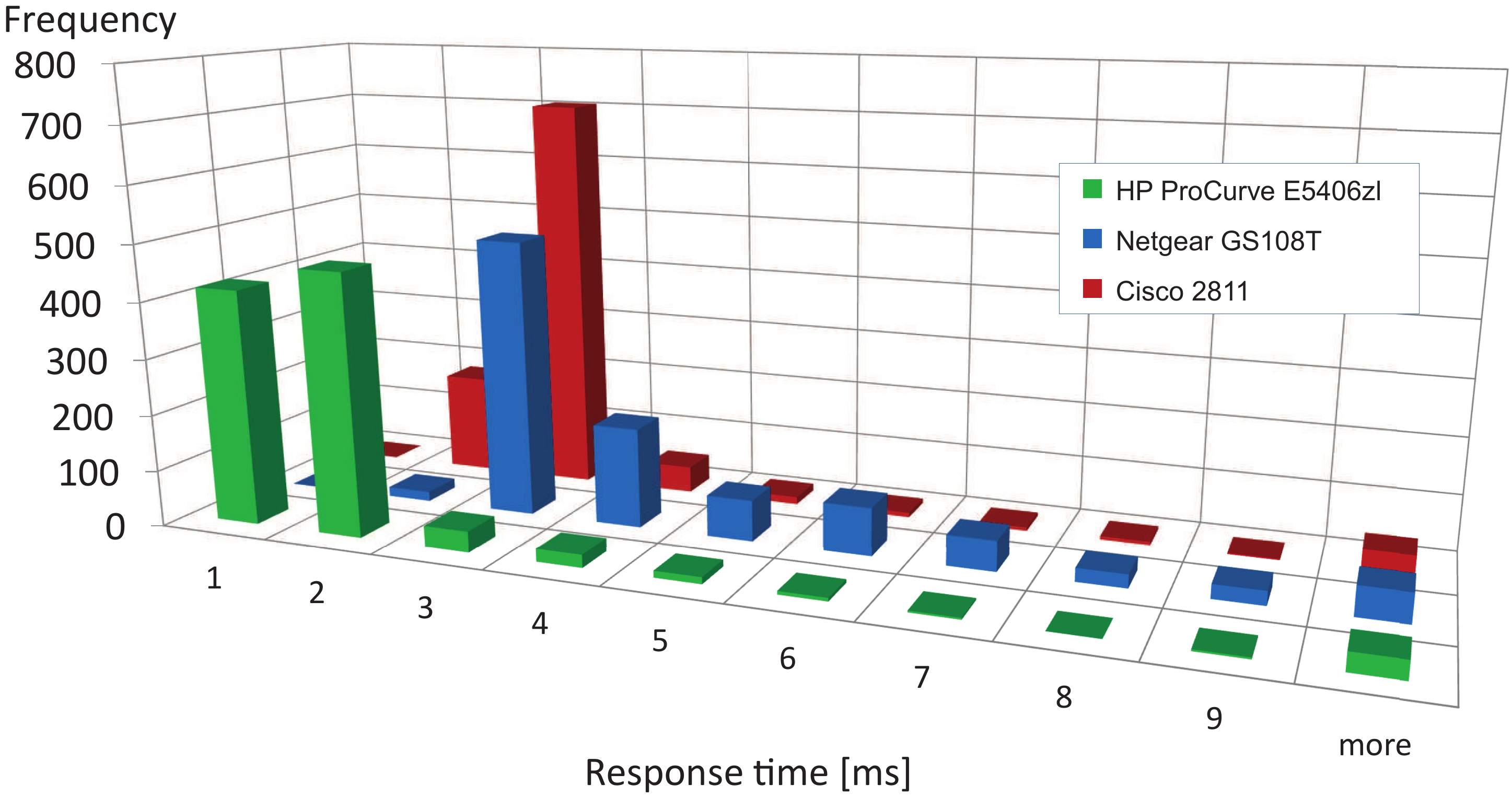}
\caption{The SNMP response time of the different devices measured in our testbed.}
\label{response_time} \end{figure}

The SNMP \emph{response} time is the time between sending a request packet and the arrival of the corresponding response packet. It includes the network transfer time and the processing time of the the device-internal SNMP agent. To understand whether SNMP queries with a relatively high frequency have impact on the packet processing performance of the equipment under load, we measured the response times of the devices in our lab, which is shown as a histogram in Figure~\ref{response_time}. The system clock of the monitoring computer served as reference time. 1000 requests were sent to each device and the figure shows that most response times are around 1-2 ms for the HP switch and mostly at 2-3 ms for the Cisco router. The Netgear switch is slightly slower and exhibits a larger variance in response times. Based on these observations, we consider all tested devices as suitable for fairly accurate throughput calculations (i.e. with measurement errors negligible for all practical purposes), assuming that the measurement time for each IE flow is an order of magnitude larger than the maximum response time of the devices.

Operators of large IP backbone networks usually have a comprehensive suit of data bases and software tools for operation, administration and maintenance for their infrastructure in place, which is commonly referred to as the Network Management System (NMS). Traffic monitoring is one of its important subsystems, which serves as source of information for many operational tasks, such as fault detection, capacity planning, anomaly analysis, etc. The monitoring subsystem required for the measurement scheme proposed in this paper must implement 1)~an interface to the SDN controller's northbound API to fetch the flow byte counters from the OpenFlow-enabled devices, 2)~an SNMP manager to fetch the link byte counters from the legacy OSPF routers, and 3)~an NMS-internal interface to the topology service subsystem (i.e. the data base that stores a model of the network topology including routing information and the specification of the resources).

\begin{figure}[t] \center
\includegraphics[width=\columnwidth]{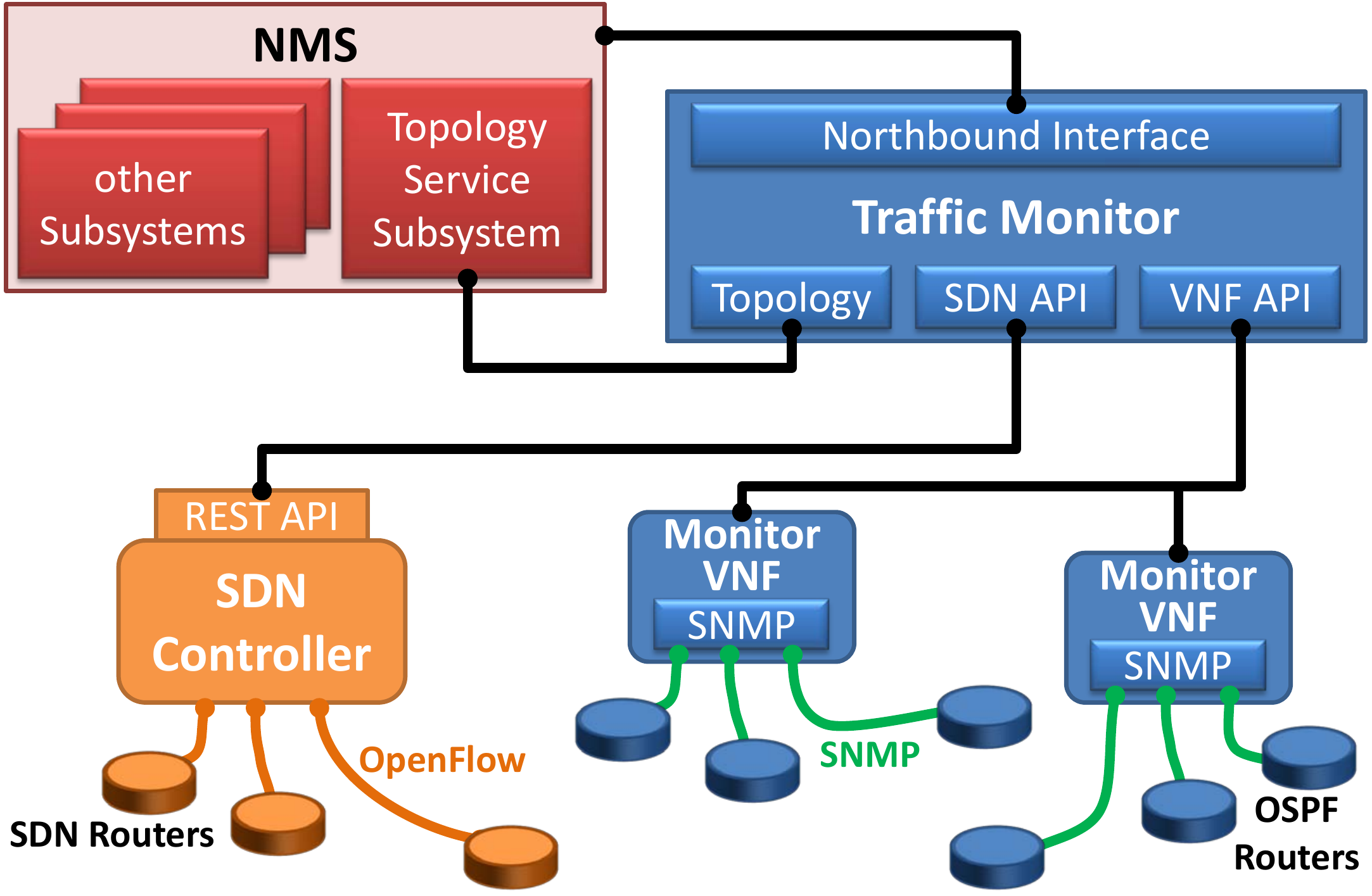}
\caption{Our software architecture implemented for distributed measurements with monitoring VNFs.}
\label{vnf} \end{figure}

As illustrated in Figure~\ref{vnf}, we implemented a simple proof-of-concept monitoring application to perform traffic measurements from link and OpenFlow byte counters, which additionally allows to temporarily reroute specific flows to isolate them for measurements on a backup link. It retrieves OpenFlow counters from the central SDN controller\footnote{We used Floodlight~\cite{floodlight} as SDN controller in our experiments, but other implementations (e.g. OpenDaylight) provide similar APIs, and could be used interchangeably.} through the controller's REST-based API. Please note that in large IP backbone topologies, the network transfer time for the SNMP counter requests, and more importantly its jitter, can become significantly large, which would affect the accuracy of the byte-counter-based measurements. We have therefore considered the deployment of the measurement application in the form of a parallelized Virtual Network Function (VNF) that can be operated in a distributed fashion, like shown in Figure~\ref{vnf}. This would allow to implement the central part of the traffic monitor (the large blue box) as a module of the NMS, whereas the actual SNMP-based measurements would be performed by lightweight measurement VNFs (that could be operated on demand in virtual machines or application containers) closer to the actual devices from where the SNMP byte counters have to be retrieved. While our proof-of-concept application has not been entirely implemented in such a modular fashion, it is planned as future architecture for this research.

\begin{figure}[t] \center
\includegraphics[width=8.7cm]{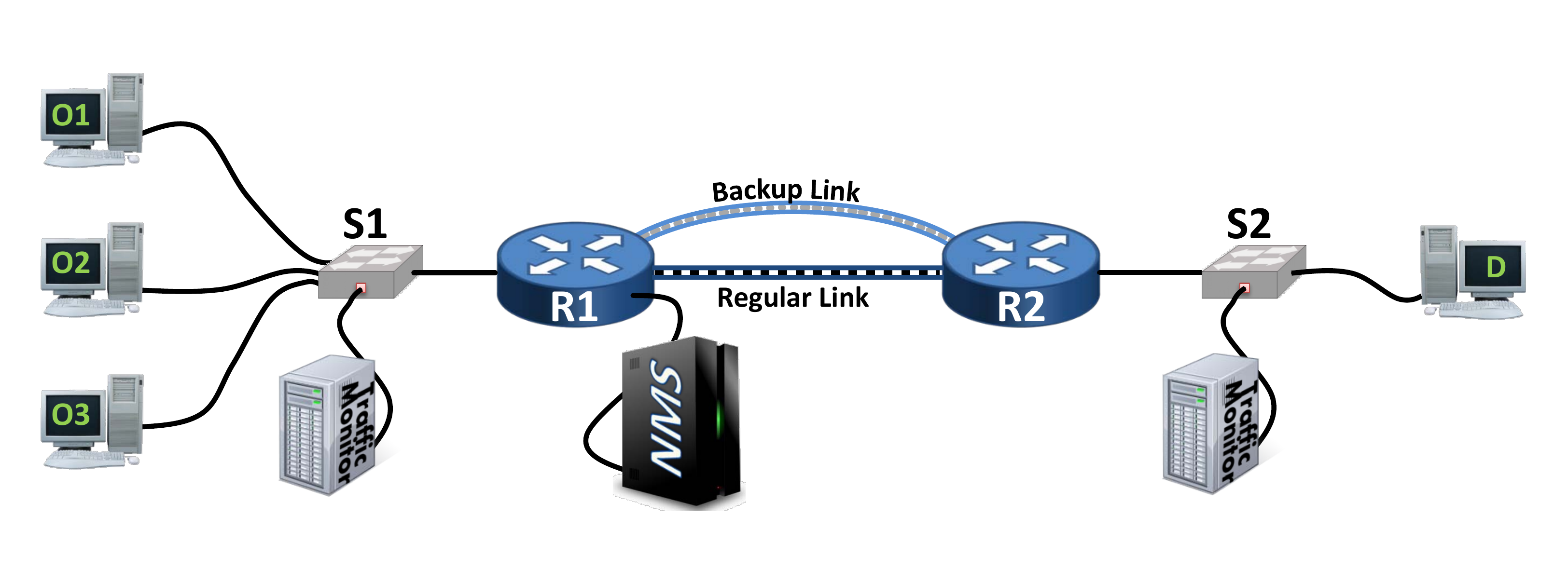}
\caption{Testbed setup for our measurements on a backup link.}
\label{testbed} \end{figure}

\subsection{Hands-on measurements}
We set up the testbed shown in Figure~\ref{testbed} to demonstrate SNMP- and OpenFlow-based measurements in our lab. The network consists of two routers, two Ethernet switches and a total of seven PCs. Four PCs represent endpoints of data connections: $O1$, $O2$, and $O3$, connected through a switch to router $R1$ serve as traffic origins, and $D$, connected to router $R2$, as destination. The routers are connected with two links: the first (i.e. working) link, denoted as backbone IP link, and the backup link. The fifth PC with the proof-of-concept NMS application is directly connected to $R1$. The incoming traffic at $R1$ and the outgoing traffic at $R2$ is measured via the intermediate switches' port mirroring function in order to obtain comparison values to the NMS measurements. We used the Iperf command line tool on the PCs to generate UDP traffic flows with constant bit rates. It should be noted that, in contrast to what is shown in Figure~\ref{arch}, our testbed does not provide dynamic and automated port configuration, as the  Layer~2 technology used is native Ethernet. In other words, the backup link setup is done manually by connecting the according devices with a patch cable. The proof-of-concept monitoring application accordingly lacks the ability to perform Layer~2 control actions.

\begin{figure}[t] \center
\includegraphics[width=\linewidth]{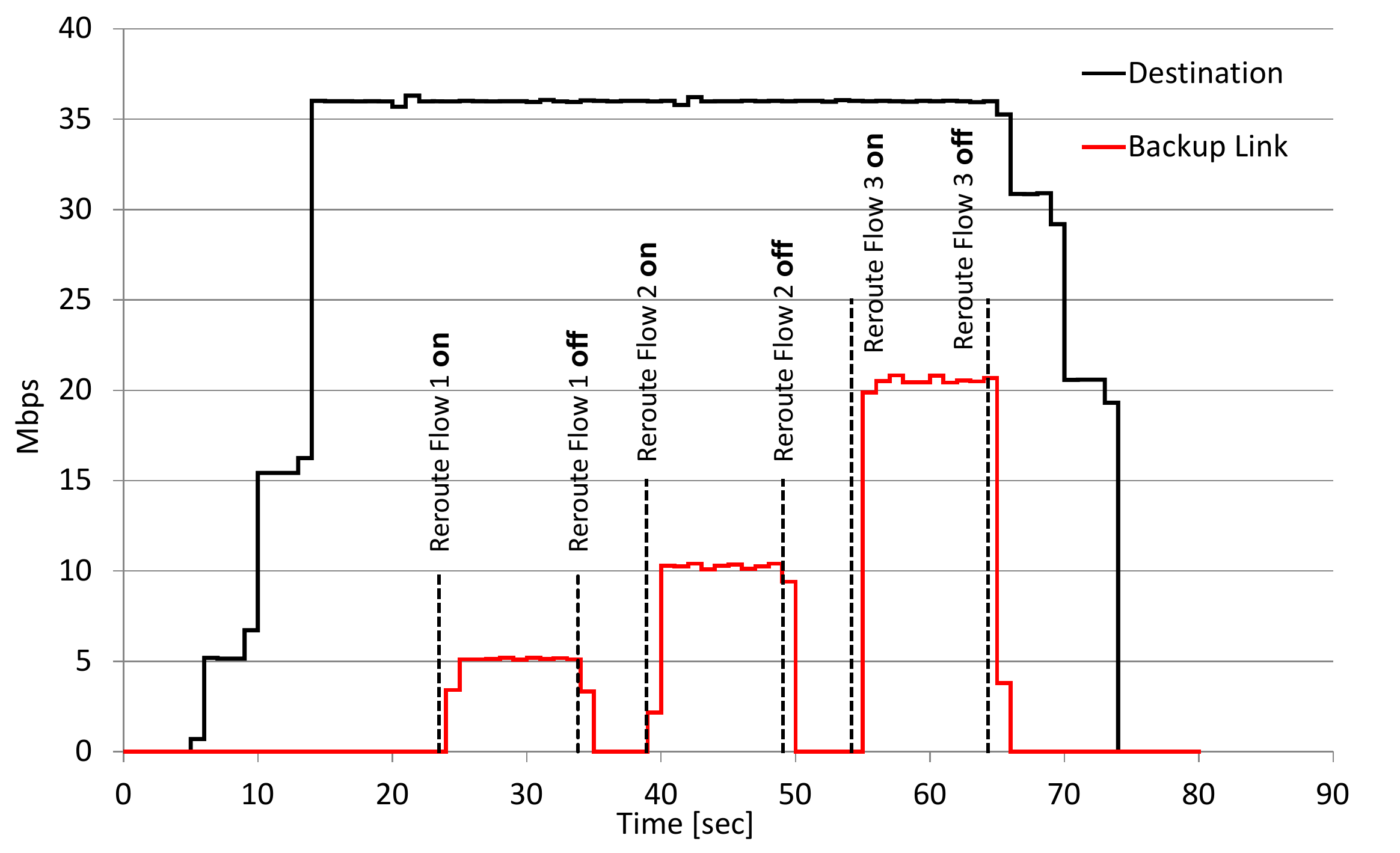}
\caption{Bit rate measurements on the backup link.}
\label{bypass_measure} \end{figure}

\begin{figure}[t] \center
\includegraphics[width=\linewidth]{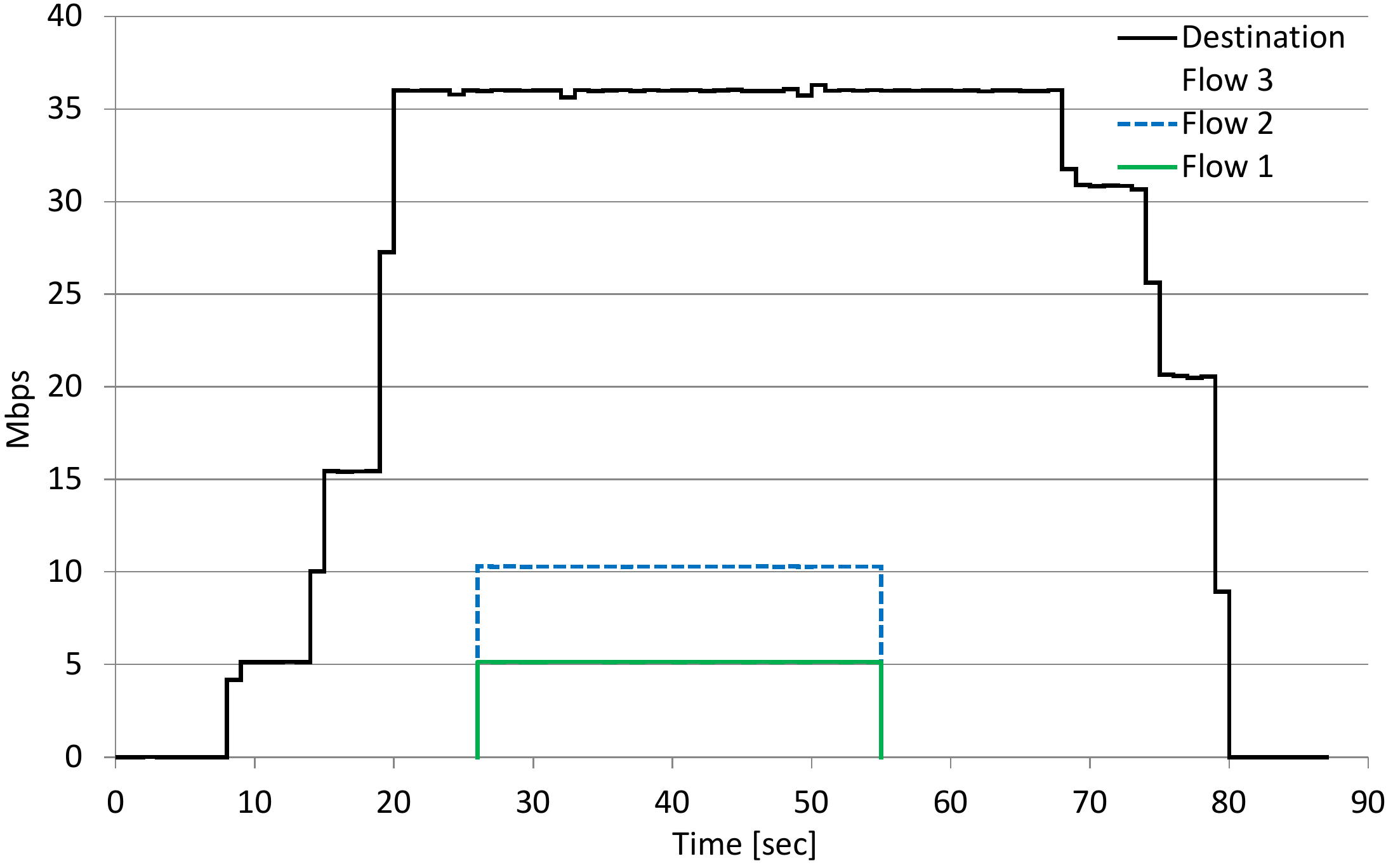}
\caption{Flow bitrate measurements with OpenFlow}
\label{openflow_measure} \end{figure}

Figure~\ref{bypass_measure} shows the measured data rates (the three bursts of the red plot) in the first experimental setup using flow separation on the backup link. The aggregated traffic was captured (along with time stamps from the system clock) with the open-source packet analyzer Wireshark at the destination host (shown as the black plot in the figure). The three IE flows add up to approximately 36~Mbit/s at the destination (including the MAC frame and IP packet headers). The individual UDP flows were configured to have 5~Mbit/s, 10~Mbit/s, and 20~Mbit/s, which, along with the L2 and L3 protocol headers\footnote{Ethernet frames have a size of 1518 bytes including the 18 bytes of the frame header (not including the Preamble and SFD) and the 20 bytes of IP packet header. Thus the overhead rate for the sent 35~Mbit/s IP payload is ca. 2.57\%, resulting in ca. 35,9~Mbit/s measured with Wireshark at the destination.} add up to 35.9~Mbit/s, like measured at the destination. The byte counter of the backup link was requested every second and the three flows were subsequently rerouted onto the backup link for ten seconds respectively. It can be seen that the measured throughput on the backup link matches the size of the three flows (plus the previously mentioned protocol overhead), which suggests that flow measurements on backup links are a practical solution to augment the monitoring system in place with additional traffic statistics.

Figure~\ref{openflow_measure} shows the measurement for the same traffic scenario in a similar testbed setup, but with an OpenFlow switch in the center instead of the two OSPF routers. We retrieved the OpenFlow byte counters for the flow table entries of the three flows (again with a time resolution of one second) through the SDN controller, and the resulting throughput of the three flows is plotted as the red, blue, and green lines. It can be seen that, like in the case of SNMP-based measurements, the values calculated from the byte counters are again matching the configured UDP data rates,  plus L2 and L3 overhead, similarly to the previous measurement.

\section{Algorithms}\label{math-section}
In order to select the most beneficial backup links and nodes for SDN deployment, and to determine which flow is measured on which device (i.e. on which backup link or on which OpenFlow router), we developed two algorithms: a)~an ILP-based, and b)~a greedy heuristic. This section presents the related optimization model and algorithm in detail. A summary of the used notation is given in Table~\ref{symbols}.

\subsection{Optimization Model}\label{ilpmodel}
The objective function of the model minimizes the total cost for the deployment of backup links and OpenFlow routers in the network:
\begin{equation}
\text{Minimize}\;\; \sum_{x\in\mathcal{L}\cup\mathcal{N}} P^x\cdot \mathsf{Cost}(x)
\end{equation}

The duration of a backup link measurement cycle (i.e., the time for retrieving the necessary SNMP link counters for all desired measurements on a particular backup link) increases with the number of sequential measurements. It is however advantageous to limit all SNMP-based measurement cycles to the duration of a predefined global monitoring interval, which allows for synchronized and complete traffic snapshots at fixed instants in time. The following constraint allows such a limitation, and it additionally limits the number of flow table entries of an OpenFlow router that can be monitored with byte counters, which is practically relevant in case the router hardware is limited in this regard.
\begin{equation}
\forall x\in\mathcal{L}\cup\mathcal{N}: \;\;\; 
\sum_{f\in\mathcal{F}} M^f_x \leq P^x \cdot \mathsf{MaxFlows}(x)
\end{equation}
The right-hand side of the constraint furthermore sets the maximum number of measurable flows to zero in case the backup link or OpenFlow switch is not deployed. In the same way, we limit the traffic load on a certain measurement device (backup link or OpenFlow switch), which may be required for one of the following reasons: 1)~An OpenFlow router's byte counter on flow table entries may be implemented only in software, so that this function is not usable at full line rate. 2)~The size of a backup link may be smaller than the size of the original link (e.g., its just a single port out of an Ethernet Link Aggregation Group), which can lead to an overutilization of the backup link when an elephant flow is measured. The following constraint can be used in such circumstances:
\begin{equation} \label{openflow-constr-1}
\forall n\in\mathcal{N}: \;\;\;
\sum_{f\in\mathcal{F}} M^f_n \cdot \hat{f }
\leq P^n \cdot \mathsf{MaxLoad}(n)
\end{equation}

\begin{table}[t]\begin{center}
\begin{tabular}{ c l }
\toprule
\textbf{Parameter} & \textbf{Meaning} \\
\midrule
$\mathcal{N}$ & Set of all nodes $n$ \\\addlinespace[1.0mm]
$\mathcal{L}$ & Set of all (directional) links $\ell$ \\\addlinespace[1.0mm]
\multirow{2}{*}{$\text{rev}(\ell) = r$} & Link reversion function for all $\ell\in \mathcal{L}$, \\
 & $\text{ }$ returns the reverse link $r$ of link $\ell$ \\\addlinespace[1.0mm]
$\mathcal{F}$  & Set of all flows $f$ \\\addlinespace[1.0mm]
$\hat{f}$ & Upper bound on the size of $f$\\\addlinespace[1.0mm]
$\mathcal{R}^f_x$ & Boolean, true if $f$ traverses $x$\\\addlinespace[1.0mm]
$\mathsf{Cost}(x)$ & Cost for the provisioning of $x$  \\\addlinespace[1.0mm]
$\mathsf{MaxFlows}(x)$ & Max number of measurable flows on $x$ \\\addlinespace[1.0mm]
$\mathsf{MaxLoad}(x)$ & Max measurable load on $x$ \\\addlinespace[1.0mm]
\midrule
\textbf{Boolean} & \multirow{2}{*}{\textbf{Meaning}} \\
\textbf{Variable} & \\
\midrule
$P^x$ & Device $x$ is provisioned \\\addlinespace[1.0mm]
$M^f_\ell$ & $f$ is measured on the backup link of $\ell$\\\addlinespace[1.0mm]
$M^f_n$ & $f$ is measured on node $n$\\\addlinespace[1.0mm]
\multirow{2}{*}{$D^f_\ell$} & $f$ is derived on link $\ell$ \\
 & $\text{ }$ from other measurements \\\addlinespace[1.0mm]
\bottomrule
\end{tabular}
\caption{Summary of Notation}\label{symbols}
\end{center}\end{table}

The next constraint assures that every flow is either measured or derived from other measurements:
\begin{equation} \label{constr1}
\forall f\in\mathcal{F}: \;\;\;
\sum_{\ell\in\mathcal{L}}\Big( M^f_\ell + D^f_\ell \Big) + \sum_{n\in\mathcal{N}} M^f_n = 1
\end{equation}

If a particular flow is the only one left undetermined on a particular link, that flow can be calculated as the total link load minus the sum of all other flows' sizes on that link. The variable $D^f_\ell$ is set to one in case the solver decides to derive a flow $f$ on link $\ell$ in this way. However, to assure that on each link at max one flow is derived in this way, we need the following constraint:
\begin{equation}
\forall \ell\in\mathcal{L}: \;\;\;
\sum_{f\in\mathcal{F}} D^f_\ell    \leq 1
\end{equation}

Apparently, a flow can only be measured on a device if it's routing path traverse the device. This is taken care of by
\begin{equation}\begin{split}
\forall f\in\mathcal{F}, \;\; \forall \ell\in\mathcal{L}, \;\; \forall n\in\mathcal{N}: \;\;\; \\
M^f_\ell \leq \mathcal{R}^f_\ell    \;\;\;\;\;\;\;\;\;    M^f_n \leq \mathcal{R}^f_n    \;\;\;\;\;\;\;\;\;    D^f_\ell \leq \mathcal{R}^f_\ell
\end{split}\end{equation}

As we consider directional links in this model, whereas IP links are bidirectional, we additionally constrain backup links to be bidirectional. In other words, if link $\ell$ is provisioned with a backup, we require that its reverse link $r$ (i.e. the one that connects the same nodes in the reverse direction) is provisioned with a backup too:
\begin{equation}
\forall \ell,r\in\mathcal{L} \;\text{with} \; r = \text{rev}(\ell) : \;\;\;
P^\ell = P^r
\end{equation}

\subsection{Greedy Heuristic}\label{heuristic}

\begin{algorithm}[t]
\KwIn{Sets $\mathcal{N}$, $\mathcal{L}$, $\mathcal{F}$, Functions $\text{trav}(x,f)$, $\text{rev}(\ell)$}
\KwOut{Set $\Omega$ of all required network resources}
\vspace{2mm}// Step 1: Initialization \\
$\Omega\leftarrow\emptyset ; \;\;\;\;$
$W\leftarrow\emptyset ; \;\;\;\;$
$R\leftarrow \mathcal{N} \cup \mathcal{L}$\\
\ForEach{$w \in R$}{
	$\vec{w}\leftarrow\vec{0}$\\
	\ForEach{$f \in \mathcal{F}$}{
		$i \leftarrow \text{ind}(f)$\\
		$\vec{w_i} \leftarrow \mathcal{R}^f_x$\\
	}
	$W\leftarrow W \cup \{\vec{w}\}$
}
\vspace{2mm}\While{$|\bigvee W| > \varphi_\text{min}$}{
\vspace{2mm}// Step 2: Determine all $W_x$ and $\varphi_x$\\
\ForEach{$x \in R$}{
	$W_x \leftarrow \emptyset$\\
	\lIf{$x\in \mathcal{L}$}{Link $r \leftarrow$ rev($x$)}\\
	\ForEach{$\vec{w} \in W\setminus\{\vec{x}, \vec{r}\}$}{
		\lIf{$x\in \mathcal{N}$}{$W_x \leftarrow W_x \cup \{ \vec{w} \wedge \neg\vec{x} \}$}\\
		\lElse{$W_x \leftarrow W_x \cup \{ (\vec{w} \wedge \neg\vec{x}) \wedge \neg\vec{r} \}$}\\
	}
	// Iteratively remove all calculable flows\\
	\While{$\exists \vec{\ell}\in W_x: \; |\vec{\ell}\,|=1$}{
		$W_x\leftarrow W_x \setminus\{\vec{\ell}\,\}$\\
		\lForEach{$\vec{w}\in W_x$}{
			$\vec{w} \leftarrow \vec{w} \wedge \neg\vec{\ell}$
		}
	}
	$\varphi_x \leftarrow |\bigvee W| - |\bigvee W_x|$\vspace{1mm}\\
}
\vspace{2mm}// Step 3: Choose next resource $x$\\
$ x \leftarrow \displaystyle\argmax_{z\in R} \{\varphi_z\} $\\
\vspace{1mm}$\Omega \leftarrow \Omega \cup \{x\}  ; \;\;\;\;$
$R\leftarrow R\setminus\{x\} ; \;\;\;\;$
$W\leftarrow W_x$
\vspace{2mm}\caption{Greedy Heuristic}\label{algo}
}
\end{algorithm}

In addition to the ILP model in the previous subsection, we provide here a greedy algorithm with its pseudocode shown in Algorithm~\ref{algo}, that can also be used as pre-stage to the ILP. This algorithm exhibits a time complexity which is orders of magnitude lower than the ILP and thus fast enough to provide solutions for large scale topologies. Its basic functioning is as follows: It determines in each iteration the next (yet not deployed) measurement resource (i.e. backup link or SDN node), whose deployment results in the largest number of determinable flows (that are yet undetermined). It therefore keeps track of which flows have been determined in previous iterations. This is necessary to avoid the case in which, for instance, there are two resources $a$ and $b$, both providing the measurement of a very high number of flows, but the majority of the flows on $a$ are the same as on $b$. A trivial greedy approach would simply choose both, while one of them would actually be redundant and without much benefit.

\begin{table}[t]\begin{center}
\begin{tabular}{ c l }
\toprule
\textbf{Vector} & \textbf{Meaning} \\
\midrule
$\vec{a}$ & Undetermined flows on network resource $a$ \\\addlinespace[1.0mm]
$\vec{a_i}$ & The $i^{th}$ element of $\vec{a}$ \\\addlinespace[1.0mm]
$\vec{0}$ & The zero vector $(0,0,\ldots,0)$ \\\addlinespace[1.0mm]
\midrule
\textbf{Connective} & \textbf{Meaning} \\
\midrule
$\neg\vec{a}$ & The element-wise negation of $\vec{a}$ \\\addlinespace[1.0mm]
$\vec{a} \wedge \vec{b}$ & The element-wise AND of $\vec{a}$ and $\vec{b}$ \\\addlinespace[1.0mm]
$\vec{a} \vee \vec{b}$ & The element-wise OR of $\vec{a}$ and $\vec{b}$ \\\addlinespace[1.0mm]
$\vec{a} \leftarrow \vec{b}$ & $\vec{a}$ is defined to be $\vec{b}$ \\\addlinespace[1.0mm]
$|\vec{a}|$ & The cardinality (i.e. number of 1-bits) of $\vec{a}$ \\\addlinespace[1.0mm]
$\bigvee A$ & The element-wise OR of all $\vec{a}\in A$ \\\addlinespace[1.0mm]
\midrule
\textbf{Function} & \textbf{Meaning} \\
\midrule
$\text{ind}(f)$ & The index number of flow $f$ in $\mathcal{F}$ \\
\bottomrule
\end{tabular}
\caption{Binary Vector Notation}\label{vectors}
\end{center}\end{table}

Our heuristic algorithm uses a working set $W$ of binary vectors (see Table~\ref{vectors} for the here introduced vector notation), which, in its initial state, represent the routing configuration in the network. Assuming that the network resources as well as the traffic flows of a network have a unique order, each vector $\vec{w}\in W$ represents a specific network resource (i.e. a node, or a directional link), and the $i^\text{th}$ element of each vector represents the same $i^\text{th}$ traffic flow. An element of a vector is 1 if the according flow traverses the according network resource, and 0 otherwise (see lines 5\ldots8 in Algorithm~\ref{algo}). The objective of the algorithm is to identify in each iteration (of the \emph{while}-loop in lines 11\ldots30) the network resource, whose deployment would result in the maximum number of new flow determinations and deletes all corresponding flows from the vectors in the working set $W$. Each iteration accordingly determines a single measurement resource (and stores it in the result set $\Omega$, see line~29) in the following fashion.

The number of flow determinations $\varphi_x$ due to the deployment of a specific resource (i.e. SDN node or backup link) $x$ is calculated (in lines 13\ldots26) as follows: We define an empty test set of vectors $W_x$ (line~14), and for each vector $\vec{w} \in W$ we add a vector $\vec{w'} \leftarrow \vec{w} \wedge \neg\vec{x}$ to $W_x$. Thus, $W_x$ represents the routing of undetermined flows after resource $x$ is deployed. Calculating $\varphi_x$ for a backup link in parallel to a link $x$ must take into account that backup links are bidirectional. In other words, the deployment of a backup link $x$ implicates the deployment of another directional link $r=\text{rev}(\ell)$ in the opposite direction. This particularity is taken care of in line~18 (which is then executed instead of line~17), where for each vector $\vec{w} \in W$ we add a vector $\vec{w'} \leftarrow (\vec{w} \wedge \neg\vec{x}) \wedge \neg\vec{r}$ to the set $W_x$.

Taking into account the \emph{calculable} flows\footnote{Like mentioned in the previous section, we can calculate the size of a particular flow, if that flow is the only one left undetermined on any (directional) link by subtracting all of the known flows on that link from its link load.} due to a resource deployment is an iterative process, because the determination of the only remaining flow on a particular link always leads on all other links (that are traversed by the said flow) to have their number of undetermined flows decreased by one. This could in turn result in another link having a single flow undetermined. We therefore iterate (lines 21\ldots24) over all link vectors $\vec{\ell}\in W_x$ with $|\vec{\ell}\,|=1$ and delete the vector from $W_x$ and then the according flow from all other vectors: $\forall \vec{w} \in W_x : \vec{w}\leftarrow \vec{w} \wedge \neg\vec{\ell}$, until there is no more such vector $\vec{\ell}\in W_x$ with $|\vec{\ell}\,|=1$. We can finally calculate the number of flow determinations (line~25) as $\varphi_x \leftarrow |\vec{z}\,|-|\vec{z_x}|$, where $\vec{z} \leftarrow \bigvee W$ and $\vec{z_x} \leftarrow \bigvee W_x$.

The final step of each iteration (line~28) is to choose the resource $x$ with the largest $\varphi_x$, and to remove all corresponding flows from the working set $W$ for the next iteration (i.e. substituting $W$ with $W_x$, line~29). The algorithm terminates when the number of remaining flows $\varphi = |\bigvee W|$ falls below a predefined threshold $\varphi_\text{min} \geq 0$ (i.e. the break condition in line~11). 

\begin{table}[t]\begin{center}
\begin{tabular}{ c c c c c }
\toprule
\textbf{Topology} & \textbf{Nodes} & \textbf{Links} & \textbf{IE Flows} & \textbf{Degree} \\
\midrule
TA2				& 65	& 108	& 4160		& 3.32 \\\addlinespace[1.0mm]
Germany50		& 50	& 88	& 2450		& 3.52 \\\addlinespace[1.0mm]
Janos-US-CA	& 39	& 61	& 1482		& 3.13 \\\addlinespace[1.0mm]
Cost266		& 37	& 57	& 1332		& 3.08 \\\addlinespace[1.0mm]
India			& 35	& 80	& 1190		& 4.57 \\\addlinespace[1.0mm]
Nobel-EU		& 28	& 41	& 756		& 2.93 \\\addlinespace[1.0mm]
France			& 25	& 45	& 600		& 3.60 \\\addlinespace[1.0mm]
New York		& 16	& 49	& 240		& 6.13 \\\addlinespace[1.0mm]
Atlanta		& 15	& 22	& 210		& 2.93 \\\addlinespace[1.0mm]
Polska			& 12	& 18	& 132		& 3.00 \\\addlinespace[1.0mm]
\bottomrule
\end{tabular}
\caption{The studied network topologies}\label{topologies}
\end{center}\end{table}

Please note that we assume in this paper the generation of the complete traffic matrix from measurements and do not take into consideration any estimation method required when the traffic matrix is not complete. However, both proposed deployment strategies in this section, i.e. the ILP-based and the heuristic, can terminate with an incomplete traffic matrix, which would require a subsequent estimation of the remaining flows. We therefore refer to the flow spread metric that was proposed in~\cite{flowspread} and represents the difference of the upper and lower bound of a flow, and thus provides a measure of \emph{urgency} for the exact determination of the flow. This flow spread value can be used as a weight metric to provide solutions that allow the measurement of the flows with the largest accumulated differences on their upper and lower bounds, and accordingly allows to minimize the estimation error.

Finally, please note that the here explained heuristic can be augmented with individual cost values for all resources. This allows a similar preference or discrimination of resources, for instance, due to reasons that
\begin{itemize}
\item specific links are more expensive to backup,
\item specific nodes are more expensive to upgrade to SDN, or
\item the upgrade to SDN of a node is in general more expensive than the backup of a link.
\end{itemize}
The number of flow determinations of a resource has then to be divided by its cost, in order to let the heuristic choose in each iteration the resource with the biggest ``return of investment''.

\section{Performance Evaluation}\label{performance-section}
In our performance evaluation, we used ten topologies from the SNDlib library~\cite{sndlib}, listed in Table~\ref{topologies}. We generated uniform distributed random values for the traffic matrices of each topology -- which has however no impact on any of the results, as we focus only on the \emph{number} of measurable and obtainable IE flows. All results were computed on an Intel Core i7-3930K CPU (6 x 3.2 GHz) and we used the GUROBI optimizer~\cite{gurobi} to solve the ILP-based problems.

\begin{figure}[t] \center
\includegraphics[width=\columnwidth]{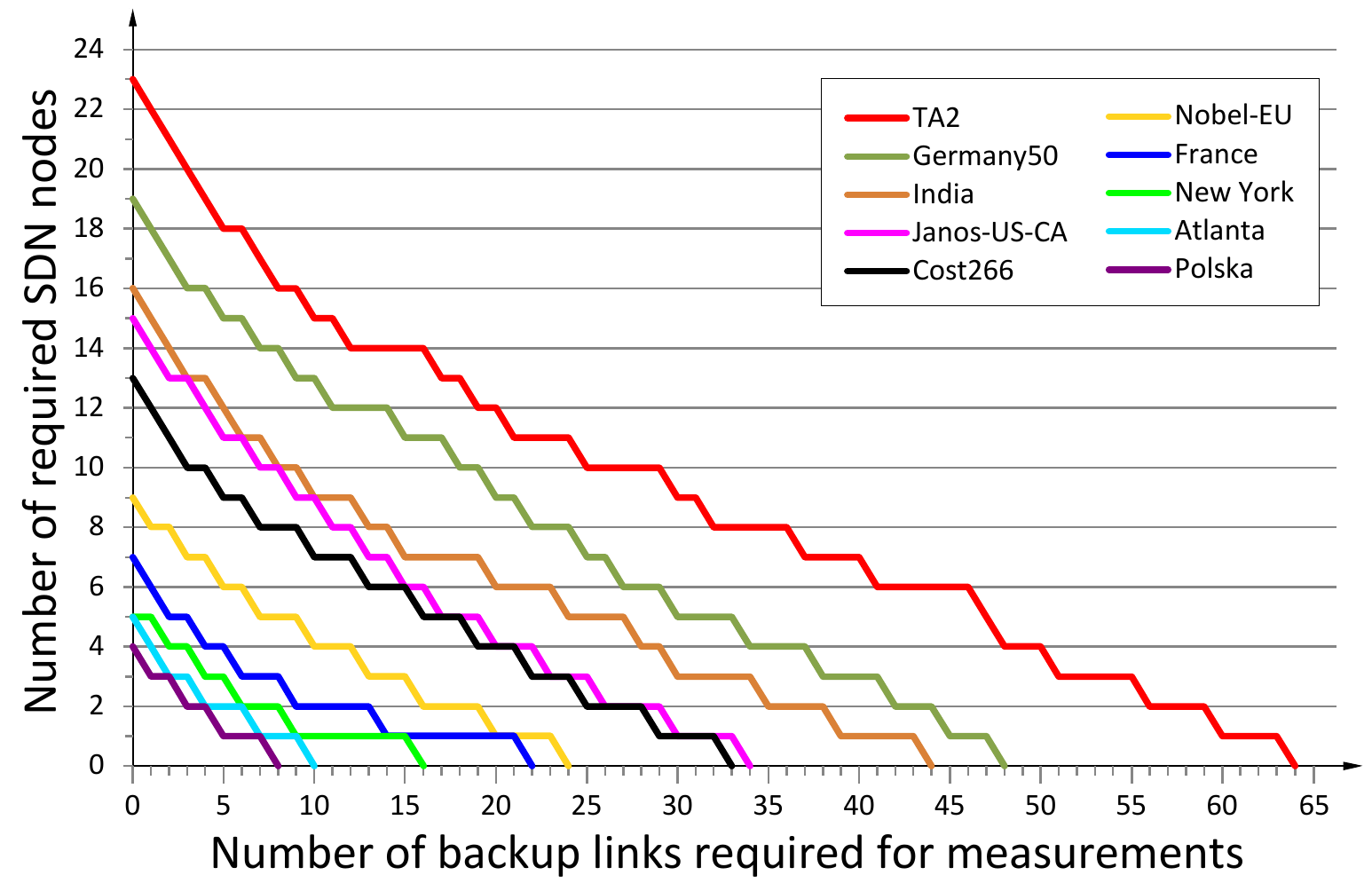}
\caption{Total number of required backup links vs. SDN nodes in various network topologies.}
\label{result_bypass_vs_sdn} \end{figure}

\subsection{SDN vs. Backup Link}
Figure~\ref{result_bypass_vs_sdn} shows our main result for the ten tested topologies: the number of required SDN nodes in the network depending on the number of deployed backup links, assuming that all the resources have been located optimal. We have used the ILP model in Subsection~\ref{ilpmodel} with an additional constraint to fix the number of backup links in order to allow the computation of the exact number of SDN nodes for any given number of backup links. Please note that the parameters MaxFlows and MaxLoad have not been used (i.e. have been set to $\infty$) in our comparison, but are provided in the model for completeness, as actual network resources may require the consideration of such limitations. It can be seen in the figure that SDN nodes are typically traversed by a larger number of flows, which results in a relatively large number of backup links if zero SDN nodes are to be used. It can also be seen that the number of required measurement resources scales with the size of the topology.

\begin{figure}[t] \center
\includegraphics[width=\columnwidth]{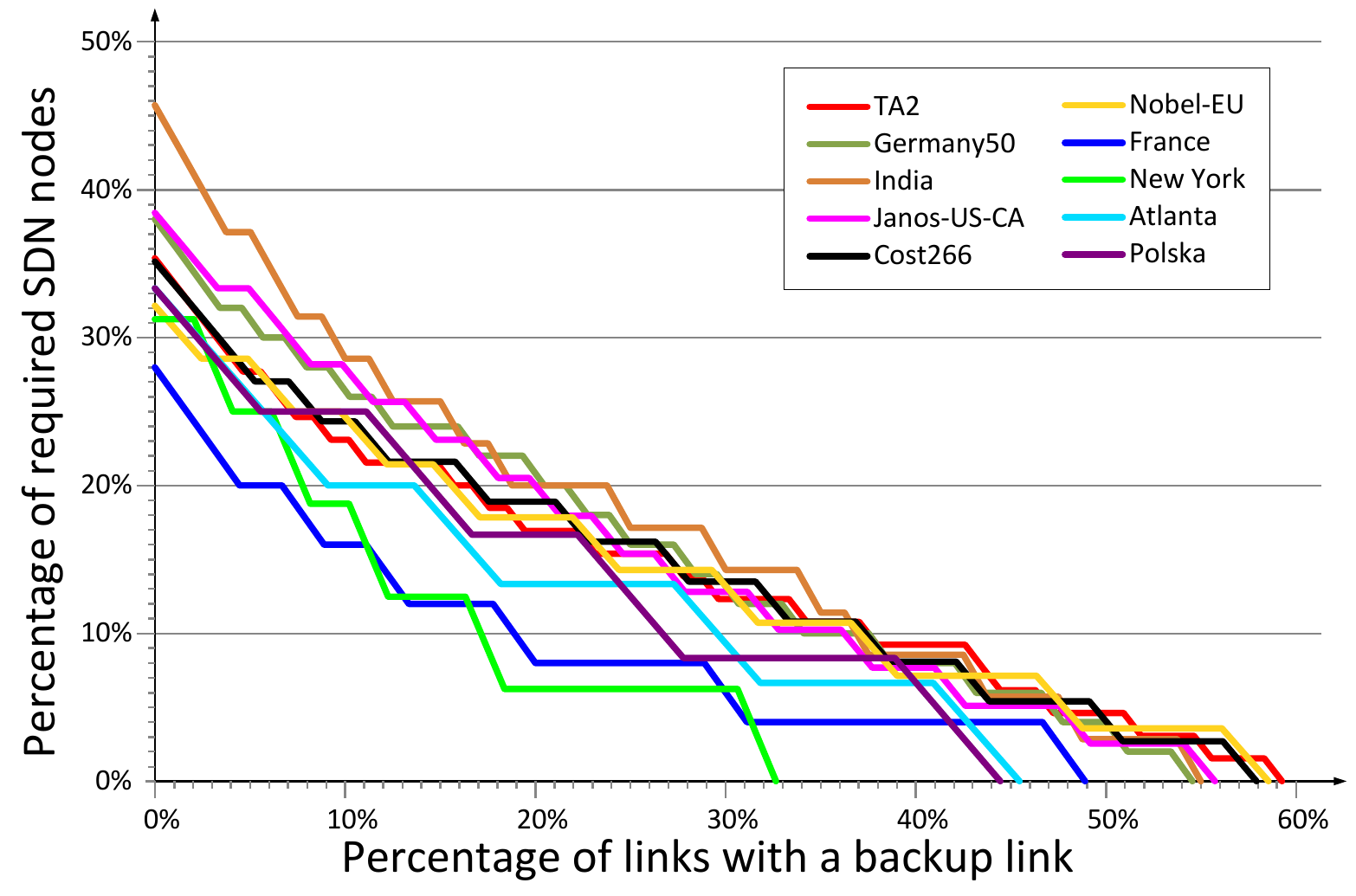}
\caption{Relative interplay of backup links and SDN node quantities in various network topologies.}
\label{result_bypass_vs_sdn_relative} \end{figure}

In contrast to the absolute numbers in Figure~\ref{result_bypass_vs_sdn}, we rescaled the plots for Figure~\ref{result_bypass_vs_sdn_relative} to provide insights independent of the network size. The figure shows that the majority of networks exhibit a similar characteristic of requiring a relatively linear combination out of 30\%-40\% SDN nodes and 50\%-60\% backup links. There are three plots that differ slightly from this pattern, which are those of the India, France and New York topologies, which are also the three networks with the largest nodal degree (see Table~\ref{topologies}) in our comparison. However, the analysis of a much larger number of topologies would be required to confirm such a principle behind the observed manner. What can however be confirmed from Figure~\ref{result_bypass_vs_sdn_relative}, is that the relation of required SDN nodes and backup links appear to be independent of the network size, as for instance the the TA2 topology and the Nobel-EU topology (which has less than half the size of TA2) exhibit a very similar characteristic.

\subsection{SDN node deployment strategies}
While the deployment of backup links by a network operator solely for the purpose of traffic measurements still appears somehow comprehensible, we assume that the deployment of SDN nodes for the same purpose is rather unrealistic due to the required cost and infrastructure upgrade effort. We have therefore tested to what extent traffic measurements can benefit from a more realistic upgrade strategy, that was proposed in~\cite{migration1}. The objective of that strategy is to provide for a given number of SDN nodes the maximum control on routing decisions to the central SDN controller, which is here measured in \emph{number of route alternatives}. It was shown in that paper that a larger total number of available paths to chose from allows for a more sophisticated traffic engineering and load balancing of the network, which appears to be a reasonable objective for network operators. 

\begin{figure}[t] \center
\includegraphics[width=\columnwidth]{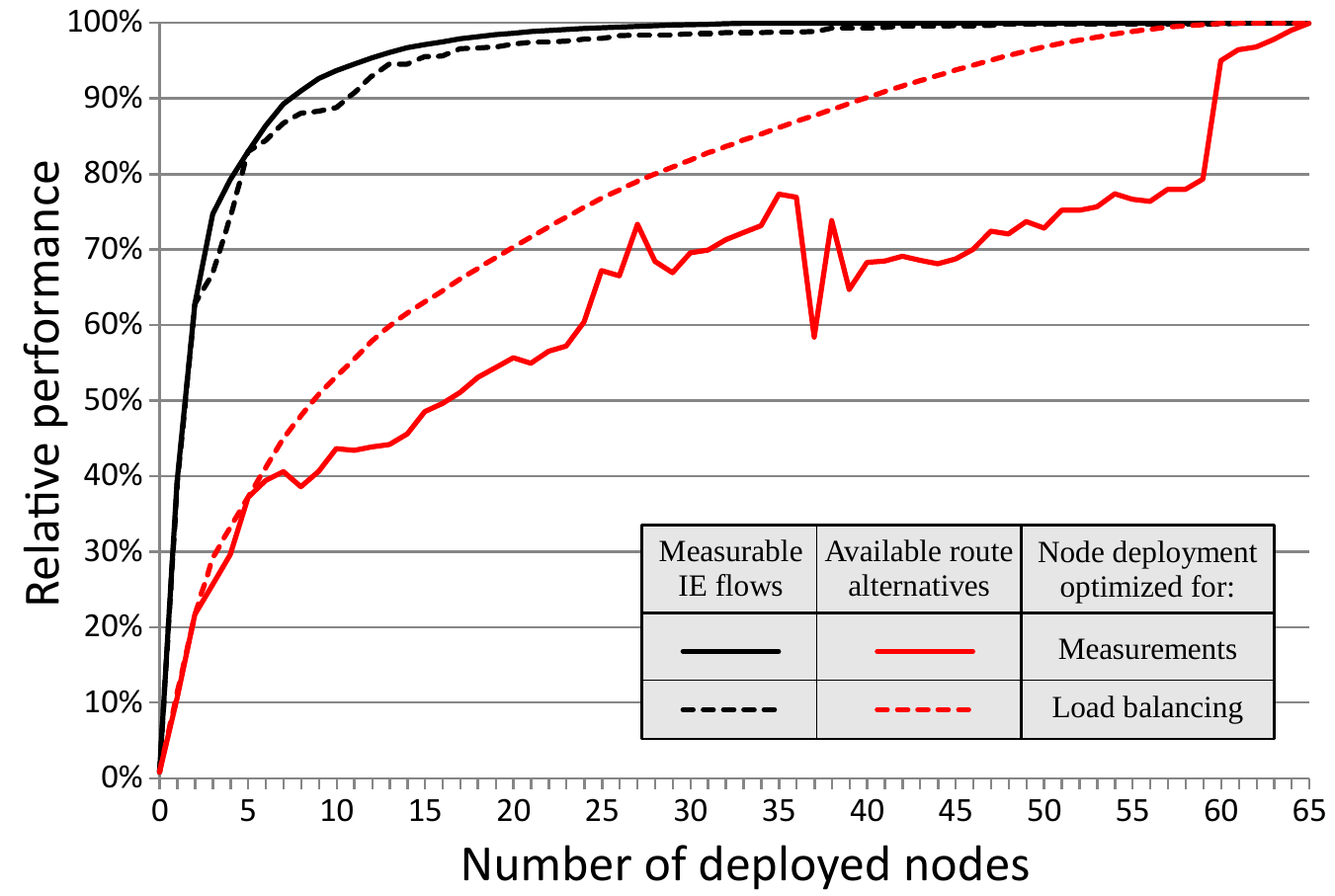}
\caption{Suitability of the load balancing node deployment strategy for traffic measurements (dashed black line) and vice versa (solid red line) in the 65 nodes TA2 topology.}
\label{comparison-result} \end{figure}

Figure~\ref{comparison-result} plots the two performance measures, i.e., number of alternative routes (the red lines) and number of measurable flows (the black lines) depending on the number of SDN nodes, using either the realistic upgrade to SDN strategy from~\cite{migration1} (the dashed lines) or the locations optimal for measurements (the solid lines). We here used the TA2 topology, which due to its size (65 nodes) provided the largest resolution of the x-axis, and we deployed solely SDN nodes (and no backup links) in order to make the plots comparable. We furthermore normalized all values with the respective maxima (i.e., 16856 alternative routes after full SDN deployment vs. a total of 4160 IE flows) and show only the relative performance on the y-axis. A comparison of the two black plots shows that the SDN upgrade strategy that maximizes routing control provides near optimal locations for measurements, as the number of measurable IE flows falls negligibly below the ones that are achievable with optimally located SDN nodes. It can furthermore be seen that the reverse (i.e. comparing the two red plots) does not hold: the node locations optimal for traffic measurements are significantly less suited for traffic engineering and load balancing. An important finding of our work is thus that operators considering to upgrade their legacy IP networks to SDN can use the strategy in~\cite{migration1} without noticeable drawbacks on SDN's traffic measurement capabilities. The chosen node locations can then be preset in the here presented ILP and heuristic to determine solely the missing backup links to complete the traffic matrix.

\begin{figure}[t] \center
\includegraphics[width=\columnwidth]{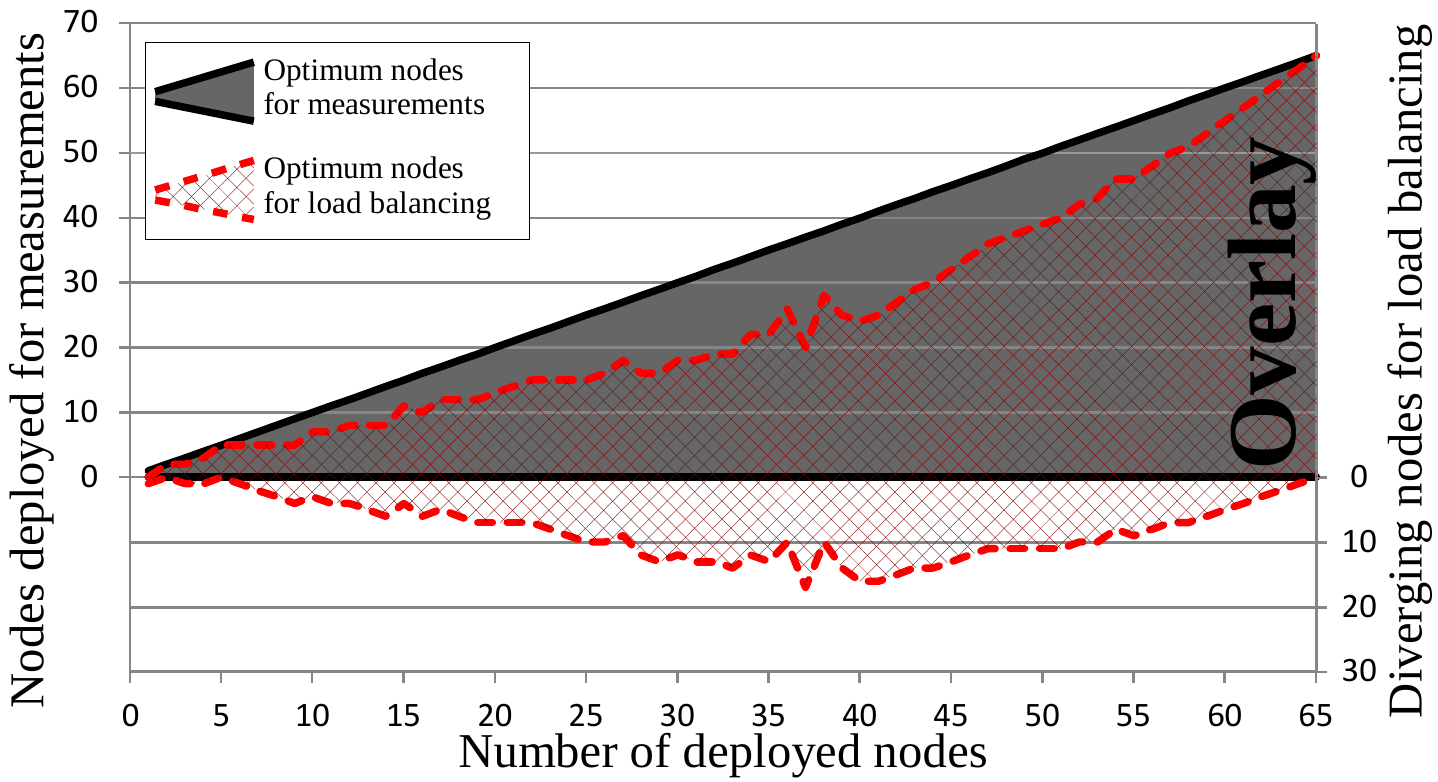}
\caption{Overlap of chosen nodes of both deployment strategies in the 65 nodes TA2 topology.}
\label{overlap-result} \end{figure}

The two strategies choose indeed very similar nodes, which we attempt to visualize in Figure~\ref{overlap-result}. The figure shows the overlap of nodes chosen from both strategies for a given number of deployable SDN nodes. The gray area depicts the optimal nodes for traffic measurements, whereas the red shaded area shows the optimum nodes for load balancing and traffic engineering. The two bounding (dashed red) lines of that area can be interpreted as following: The upper line plots how many of the nodes optimally deployed for load balancing are also optimal for measurements (left y-axis). The lower line plots the number of nodes optimally deployed for load balancing that have not been chosen by our measurement location optimization (right y-axis).

\subsection{Performance of the heuristic algorithm}
The heuristic algorithm in Subsection~\ref{heuristic} can be used instead of the linear optimization model in Subsection~\ref{ilpmodel} in case finding the optimal solution exceeds acceptable computation times due to the network's size. We propose for accuracy to use the heuristic only for a subset of the required resources (i.e., to terminate the heuristic before the solution is complete), and to preconfigure the ILP model with the chosen resources to find the remaining resources.

\begin{figure}[htb] \center
\includegraphics[width=\columnwidth]{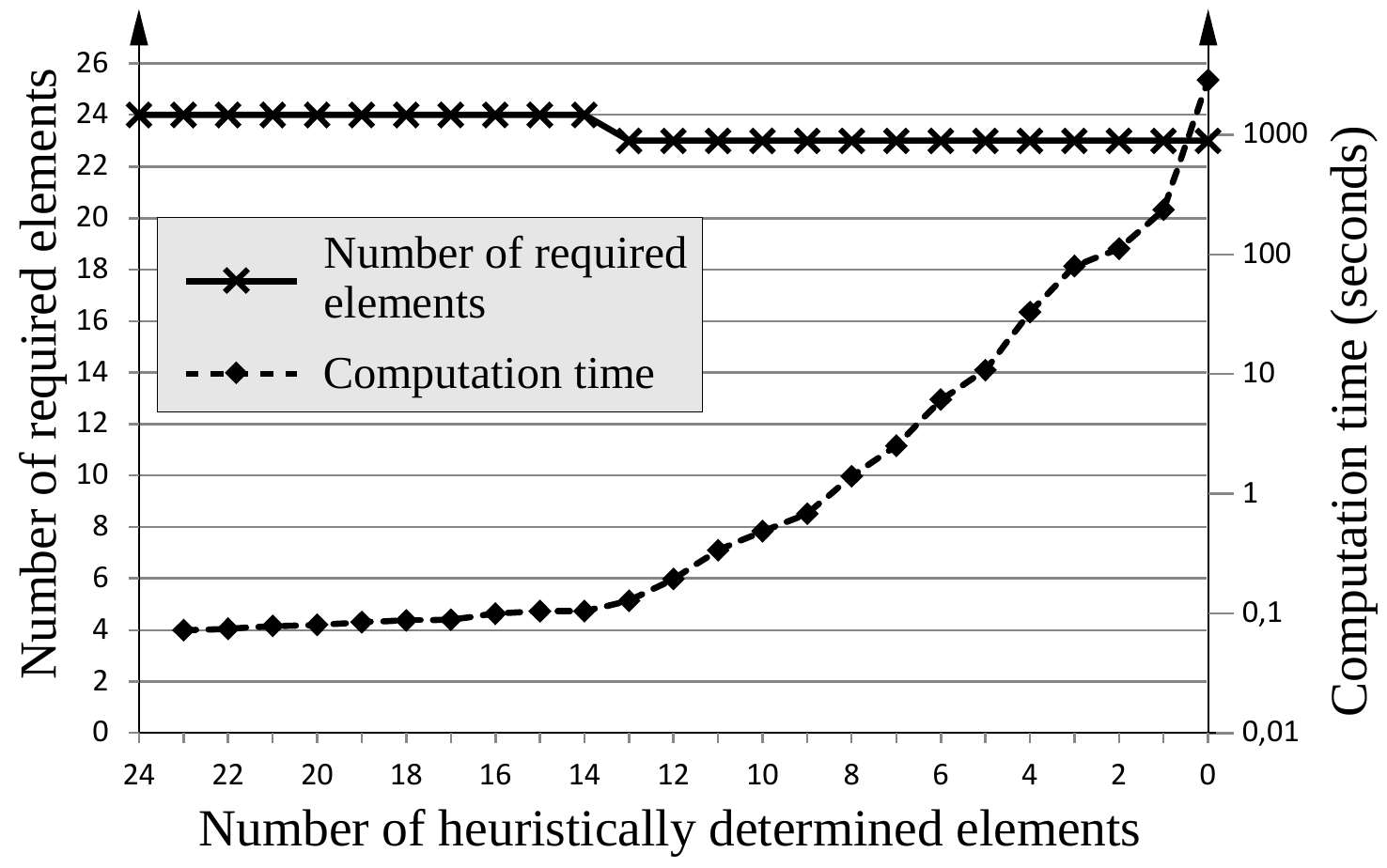}
\caption{Quality of the greedy algorithm depending on the number of chosen elements in the 65 nodes TA2 topology.}
\label{heuristic2-result} \end{figure}

\begin{figure}[t] \center
\includegraphics[width=\columnwidth]{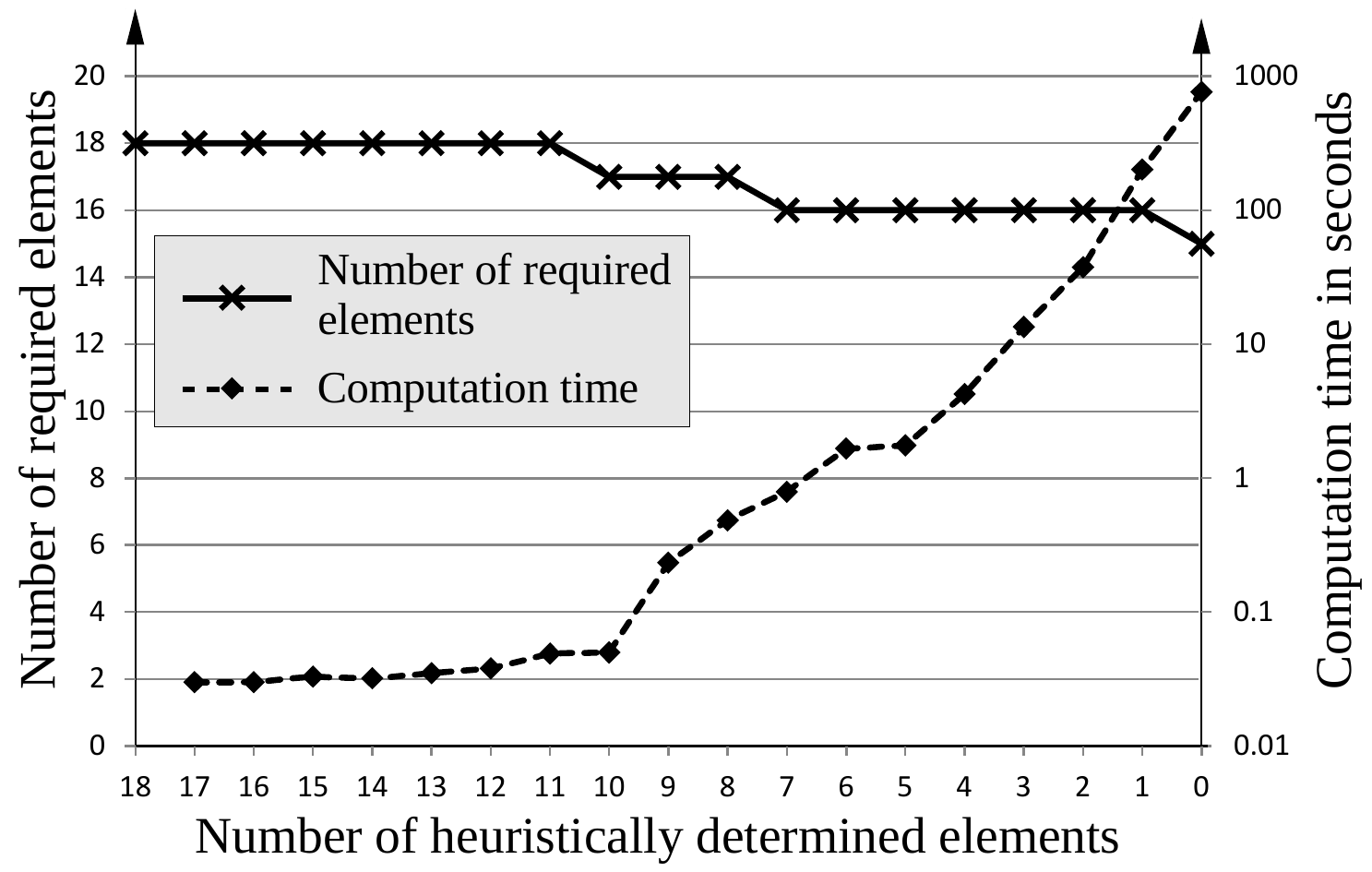}
\caption{Quality of the greedy algorithm depending on the number of chosen elements in the 39 nodes Janos-US-CA topology.}
\label{heuristic-result} \end{figure}

Figures~\ref{heuristic2-result} and~\ref{heuristic-result} show the total number of required resources (solid line, left y-axis), depending on to what extent the problem was solved with the heuristics (x-axis), before the remaining resources were determined with the ILP model. The less resources are heuristically determined, the larger is the time complexity of the remaining linear optimization problem, which can be observed in the second plot (dashed line, right y-axix) in both figures: While the heuristic (compared to the ILP) terminates in negligible time\footnote{The required computation time of our heuristic was below five seconds in a 1000 node random topology that we additionally tested.}, the time to find the optimal locations of the remaining resources increases beyond exponentially with the number of those resources.

We show each of the two results in this subsection for different purposes: Figure~\ref{heuristic2-result} shows the discussed behavior for the TA2 topology, which is the largest out of the ten compared ones, and thus the most demanding in terms of the optimality. Especially the last four data points of the time plot show that the time complexity becomes prohibitive large. It should be noted that the initialization of the optimization model requires a fixed duration depending on the network size and independent of the actual problem size, which is why the time values in the x-axis range between 23 and 14 preset resources appear to be somewhat constant. The 65 nodes of the TA2 topology can therefore be considered as borderline tractable regarding time complexity for the ILP.

Figure~\ref{heuristic-result} shows the same result for the Janos-US-CA topology, which we chose because we observed the comparably worst performance of the heuristic amongst the tested topologies: The heuristic alone determines 18 resources for the complete traffic matrix (the leftmost data point), whereas the optimal solution requires only 15 resources (the rightmost data point). This suggests that the heuristic should only be used to the point where the search for the remaining resources by an ILP solver is acceptable, for instance, by iteratively reducing the number of heuristically preconfigured resources.


\section{Conclusions}\label{conclusion-section}
This paper examined to what extent hybrid SDN/OSPF can solve the IP traffic matrix and related monitoring problem, inherent to the IP layer. To this end, we proposed a novel approach to generate the IP traffic matrix from measurements of individual ingress-egress flows using both types of byte counters, from backup links between legacy routers and flow table entries of OpenFlow-enabled routers. Instead of using expensive monitoring infrastructure for non-SDN devices, we propose to use policy based routing for backup ports and SNMP-based byte counters, features that are likely to be readily available in IP networks. We showed that our method does not impact the IP routing in place, detailed the necessary configurational steps at the backup link ingress, and discussed SNMP timing issues. 

\par We also presented a software architecture for parallelized traffic measurements based on distributed VNFs that are connected to a central monitor, which allows to prevent timing-related measurement errors due to long transmission times in large network topologies. The experiences we made with our proof-of-concept implementation in our testbed confirm the applicability of our approach even in a networking environment containing outdated equipment. We finally provided a linear optimization model and a heuristic algorithm for combined SDN node and backup link placement that assures the retrieval of the full traffic matrix under minimum resource requirements.

Our numerical evaluation showed that there is a near linear trade-off between SDN nodes and backup links that are required for a full traffic matrix, which lets us conclude that a hybrid network with a few SDN nodes can already provide complete traffic statistics, when enough backup links are available for SNMP-based measurements. We have finally shown as one of our main result in our analysis that the proposed SDN deployment strategy for traffic measurements in hybrid networks is very compatible with SDN upgrade strategies that aim for maximum network control.

\section*{Acknowledgments}
This work has been supported by the German Federal Ministry of Education and Research (BMBF) under the EUREKA / Celtic-Plus project SENDATE-PLANETS.

\vspace{2mm} We thank our student Alexander Hoffmann for the implementation of our proof-of-concept measurement application and for carrying out the measurements of MIB update rates, SNMP response times, and the SNMP- and OpenFlow-based throughput.

\end{document}